\documentclass[twocolumn]{aastex631}
\usepackage{amsmath}
\usepackage{graphicx}
\DeclareGraphicsExtensions{.pdf,.eps,.png}

\shorttitle{Jet-cocoon for EP240414a}
\shortauthors{Zheng et al.}

\begin{document}

\title{EP240414a: Off-axis View of a Jet-Cocoon System from an Expanded Progenitor Star}

\author[0000-0001-5751-633X]{Jian-He Zheng}
\affiliation{School of Astronomy and Space Science, Nanjing University, Nanjing 210023, People’s Republic of China}
\affiliation{Departments of Astronomy and Theoretical Astrophysics Center, UC Berkeley, Berkeley, CA 94720, USA}

\author[0000-0002-9195-4904]{Jin-Ping Zhu}
\affiliation{School of Physics and Astronomy, Monash University, Clayton Victoria 3800, Australia}
\affiliation{OzGrav: The ARC Centre of Excellence for Gravitational Wave Discovery, Australia}

\author[0000-0002-1568-7461]{Wenbin Lu}
\affiliation{Departments of Astronomy and Theoretical Astrophysics Center, UC Berkeley, Berkeley, CA 94720, USA}

\author[0000-0002-9725-2524]{Bing Zhang}
\affiliation{Nevada Center for Astrophysics, University of Nevada Las Vegas, Las Vegas, NV 89154, USA}
\affiliation{Department of Physics and Astronomy, University of Nevada Las Vegas, Las Vegas, NV 89154, USA}


\begin{abstract}
When a relativistic jet is launched following the core-collapse of a star, its interaction with the stellar envelope leads to the formation of a hot cocoon, which produces various viewing-angle-dependent observational phenomena following the breakout from the surface. We study the observational signatures of fast X-ray transient (FXT) EP240414a, which may originate from a jet-cocoon system viewed slightly off-axis. 
In our model, (1) the prompt X-ray emission lasting $\sim\! 100\,{\rm{s}}$ is attributed to the cooling emission from the inner cocoon (shocked jet material);
(2) the $\sim\! 0.1\,{\rm{d}}$ X-ray emission comes from the inner cocoon's afterglow; (3) the $\sim\! 0.4\,{\rm{d}}$ thermal-dominated optical emission arises from the cooling of the outer cocoon (shocked stellar material); (4) the $\sim\! 3\,{\rm{d}}$ non-thermal optical component and subsequent radio emission can be explained by the afterglow from a jet with a viewing angle of $10^{\circ}\lesssim \theta_{\rm{v}}\lesssim15^\circ$; and (5) the associated broad-lined Type Ic supernova only dominates the optical emission after $\sim\! 7\rm\, d$.
Both the jet inferred from the off-axis afterglow and the inner cocoon constrained by the cooling emission are found to have similar kinetic energies, on the order of $10^{51}\,{\rm{erg}}$. We find that the progenitor's radius is $\sim3\,R_\odot$ as constrained by the { inner cocoon's} cooling emissions, indicating that the pre-explosion star may be a massive helium star that is slightly inflated. More FXTs associated with off-axis jets and supernovae will be further examined by the Einstein Probe, leading to a deeper understanding of jet-cocoon systems.

\end{abstract}


\section{Introduction} 
Massive stars undergo core collapse at the end of their lives, potentially producing luminous explosions such as supernovae (SNe) and long-duration gamma-ray bursts (LGRBs) \citep{Woosley2006ARA&A..44..507W, KumarZhang15, Zhang2018pgrb.book.....Z}. 
The association of LGRBs with SNe \citep[e.g., GRB980425/SN\,1998bw;][]{Galama1998Natur.395..670G,Woosley1999ApJ...516..788W} provides the strongest support for the collapsar model, in which relativistic jets can be launched from a central engine embedded within the progenitor \citep{Woosley1993,MacFadyen1999,Zhang2003, fujibayashi23_collapsar, Shibata25_collapsar}. 

When a relativistic jet propagates through the stellar envelope, a hot cocoon that surrounds the jet can be formed \citep[e.g.][]{RamirezRuiz02, Matzner2003MNRAS.345..575M,Zhang2003,Bromberg2011ApJ...740..100B,bromberg16_MHD_jet,Nakar2017ApJ...834...28N,DeColle2022MNRAS.512.3627D}. 
After breaking out from its progenitors, the cocoon can produce various observational phenomena, including non-thermal afterglows from radio to X-rays and thermal cooling emission \citep{Nakar2017ApJ...834...28N, DeColle2018ApJ...863...32D,DeColle2018MNRAS.478.4553D}. However, these signals are rarely identified in LGRBs because they are overshadowed by the bright prompt gamma-ray emission and afterglow along the line of sight near the jet axis \citep[within a few degrees,][]{Beniamini2019MNRAS.482.5430B,Connor2024MNRAS.533.1629O}. 

In a few broad-lined Type Ic supernovae (SNe Ic-BL) such as GRB\,171205A/SN\,2017iuk \citep{Izzo2019Natur.565..324I} and SN\,2020bvc \citep{Ho2020ApJ...902...86H,Izzo2020A&A...639L..11I}, an early optical excess along with high-velocity material has been interpreted as thermal cocoon cooling.  
Moreover, \cite{Ho2022ApJ...938...85H} reported several plausible candidates for afterglows from ``dirty fireballs,'' which likely originate from jet-driven cocoons \footnote{Dirty fireballs refer to a class of hypothesized ejecta with small initial Lorentz factors $\Gamma_0\ll100$ due to significant baryon loading, which do not produce GRBs \citep{Huang2002MNRAS.332..735H,Rhoads2003ApJ...591.1097R}.}. 

The recent launch of the Einstein Probe (EP) has significantly enhanced the cadence and detectability of X-ray sky surveys, thanks to its large wide field, high sensitivity, and all-sky monitoring capability \citep{Yuan2022hxga.book...86Y}. 
Since beginning its operation in 2024, EP has detected numerous fast X-ray transients (FXTs).  
Due to the challenges of follow-up and joint observations, most FXTs detected by EP are ``orphan'' X-ray sources, making their origins difficult to identify, while some have been reported to be associated with GRBs \citep[e.g., EP240315a,][]{Liu2024arXiv240416425L,Levan2024arXiv240416350L}. 

Recently, EP240414a, with prompt X-ray emission lasting for $\sim\!110\,{\rm s}$ (observer's frame) and peak luminosity of $\sim\!10^{48}\,{\rm erg\,s^{-1}}$, was detected by the wide-field X-ray telescope (EP-WXT) onboard EP. The X-ray prompt emission evolved with a very soft energy spectrum and low peak energy $E_{\rm p}<1.3\,{\rm keV}$, distinguishing it from known LGRBs, X-ray flashes, and low-luminosity GRBs \citep{Sun2024arXiv241002315S,vanDalen2024arXiv240919056V}. 
Subsequently, its optical counterpart was discovered at a redshift of $z=0.401$, with the lightcurve displaying at least four distinct components. Initially, the optical emission is non-thermal at $\sim\!0.1$\,d (hereafter host rest frame), and then the spectrum evolved during $0.4$ to $1$\,d to be thermal-dominated, exhibiting a color temperature similar to those of fast blue optical transients near their peak luminosities.
More surprisingly, the lightcurve rebrightened and peaked at $\sim\!3\,{\rm d}$, with the color temperature and peak flux being inconsistent with a thermal origin \citep{Srivastav2024arXiv240919070S}. A spectroscopically confirmed SN Ic-BL, namely SN\,2024gsa, was then observed, whose features are similar to those of classic LGRB-associated SNe \citep{Sun2024arXiv241002315S,Srivastav2024arXiv240919070S,vanDalen2024arXiv240919056V}. 

Although no gamma-ray counterparts were reported at the time of the detection, the properties of the late-time radio emission from EP240414a were found to be consistent with those of LGRBs, suggesting it could be produced by an off-axis or a low-luminosity jet \citep{Bright2024arXiv240919055B}. Theoretical interpretations for such complex multi-wavelength lightcurve evolution and the origin of this peculiar FXT EP240414a are still unclear. For an off-axis viewing angle, due to the lack of bright on-axis jet emissions, it is expected that the early cocoon emission could be observed and used to constrain the properties of the progenitor star. 

In this paper, we present a self-consistent picture to explain the nature of EP240414 as a jet-cocoon system viewed from off-axis. { In the preparation of the current manuscript, we noticed another independent work by \citet{Hamidani2025_2025arXiv250316243H} who proposed a different picture based on a low-luminosity (marginally choked) jet viewed on-axis. Their model can also reasonably explain the rich phenomena of this transient, except that they predict a less sharply rising and decaying lightcurve in the optical band near $\sim\!3\,{\rm d}$ (our Phase 3) than our model. }
The paper is organized as follows: In Section \ref{sec:phy}, we provide an overview of the jet-cocoon system's physics. Then, we apply this physical model to explain the observational properties of EP240414a in Section \ref{sec:app}. The summary and discussion are in Section \ref{sec:Summary}.
Hereafter, a standard $\Lambda$CDM cosmology with $H_0=70\,{\rm km\,s^{-1}\,Mpc^{-1}}$, $\Omega_{\rm m}=0.3$, and $\Omega_\Lambda=0.7$ is adopted. All times and luminosities are referred to in the local rest frame of the host galaxy. We use $Q_x\equiv Q/10^x$ in cgs units.

\section{Physical Properties of Jet-Cocoon System}
\label{sec:phy}
\begin{figure}
    \centering
    \includegraphics[width=1.0\linewidth]{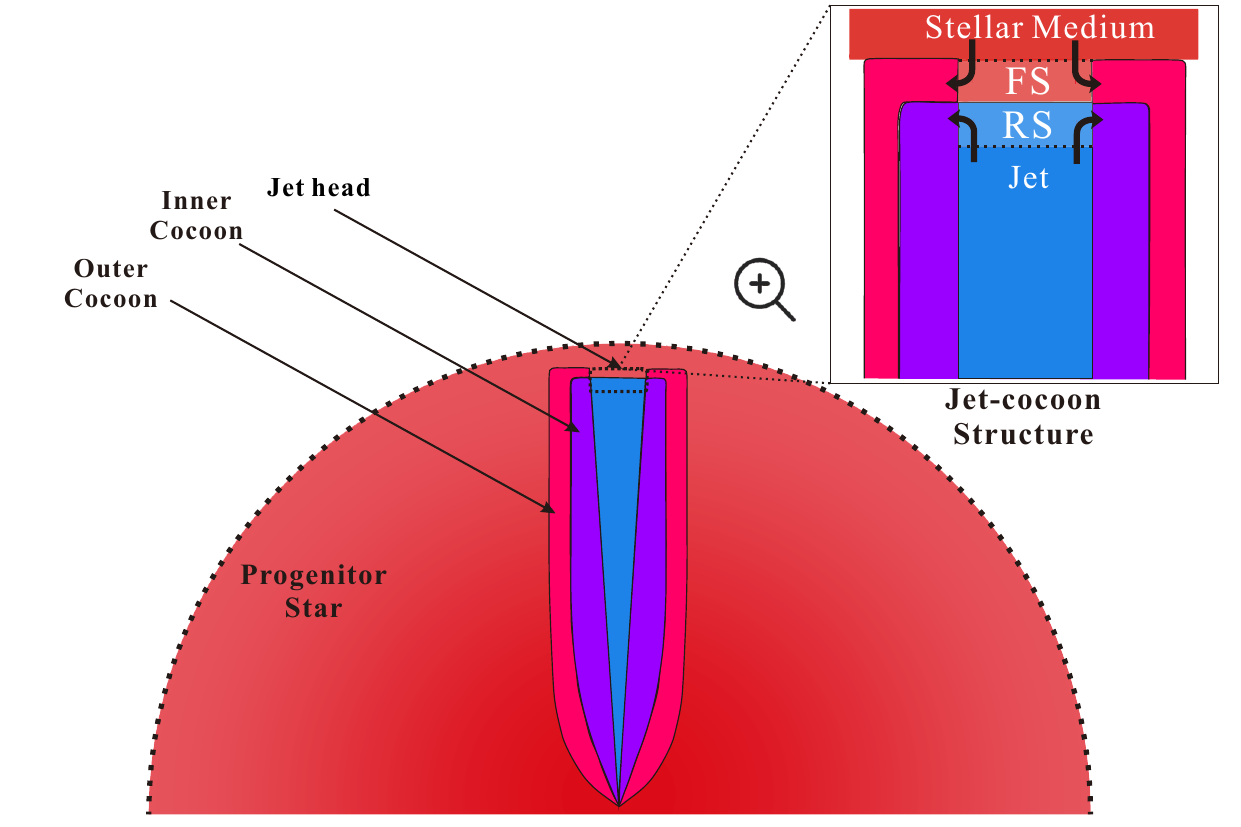}
    \includegraphics[width=1.0\linewidth]{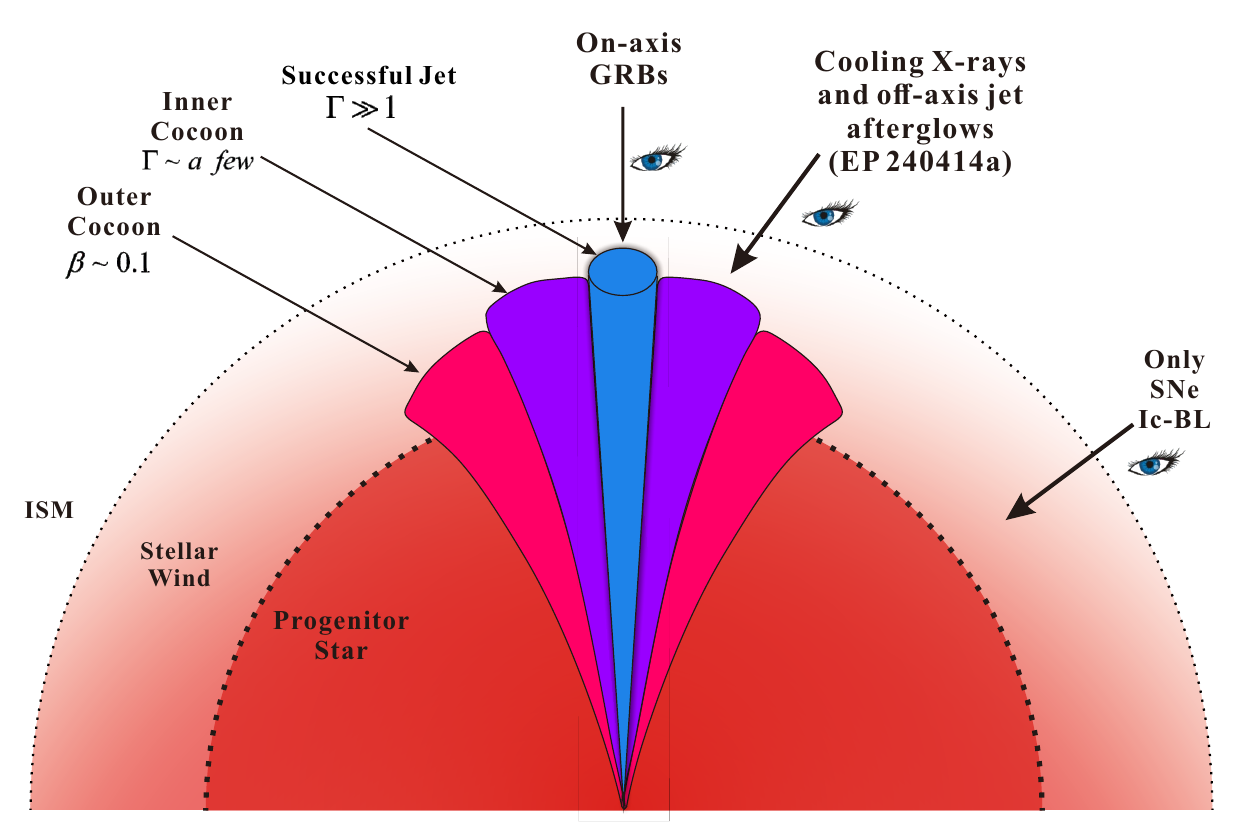}
    \caption{The schematic picture of the jet-cocoon system. \textit{Upper panel:} Before the breakout, the jet penetrates through the stellar envelope, forming a forward-reverse shock structure at the jet head. The materials swept up by the shocks are heated and flow sideways, forming two cocoon components surrounding the jet. The jet head structure is zoomed in and shown in the panel's top-right corner, with the flow directions marked by the black arrows.
    \textit{Lower Panel:} After the breakout, the relativistic jet 
    produces bright gamma-ray emission, observable by the observers within a few degrees from the jet axis (``on-axis''). The inner cocoon is mildly relativistic with early emission detectable for viewing angles $\theta_{\rm v}\sim 10^{\circ}$. The outer cocoon is non-relativistic and the emission is more isotropic. At larger viewing angles, the observable emissions are mainly contributed by the outer cocoon and the SN Ic-BL. 
    }
    \label{cartoon}
\end{figure}

As the jet propagates through the progenitor's envelope, 
their collision can lead to a forward shock (FS) sweeping up the stellar material and a reverse shock (RS) accumulating jet material at the location of the ``jet head'' \citep[e.g.][]{Matzner2003MNRAS.345..575M,Bromberg2011ApJ...740..100B}. The jet can dissipate its kinetic energy into the shock-heated gas, which flows sideways to produce a pressurized cocoon surrounding the jet (see the upper panel of Figure \ref{cartoon}). The cocoon has two components: an inner cocoon (shocked jet material) and an outer cocoon (shocked stellar material), the latter having a greater mass and a lower expansion velocity. 

After the jet successfully breaks out, from the polar axis outward (see the lower panel of Figure \ref{cartoon} for structure), there are the ultra-relativistic jet (an initial Lorentz factor of $\Gamma_{\rm j0} \sim 100$ to 1000), the typically mildly relativistic inner cocoon ($\Gamma_{\rm ic0} \sim \text{a few to 10}$), and the non-relativistic outer cocoon (a dimensionless velocity of $\beta_{\rm oc}$ of the order $0.1$). The typical initial opening angles for the jet, inner cocoon, and outer cocoon are $\theta_{\rm j0}\sim3^\circ$, $\theta_{\rm ic0}\sim15^\circ$, and $\theta_{\rm oc0}\sim45^\circ$, respectively \citep[e.g., ][]{DeColle2018MNRAS.478.4553D,Gottlieb2021MNRAS.500.3511G}. Unlike the jet and inner cocoon, whose emissions are highly viewing-angle-dependent due to their relativistic motions, the outer cocoon is quasi-spherical due to its slower velocity and its emission can be treated as isotropic. 
Hereafter, we note that the subscripts ``j,'' ``ic,'' ``oc,'' and ``c'' represent the jet, inner cocoon, outer cocoon, and one of the cocoon components (agnostic), respectively; and the subscript ``0'' means the initial value.

Significant internal energy is deposited inside the cocoons through shock heating. The total energy of the cocoons is expected to be comparable to the jet energy, typically on the order of $10^{51}$\,erg \citep[e.g.][]{Cenko2010ApJ...711..641C,Shivvers2011ApJ...734...58S}. The cocoons are expected to have two emission components: thermal photons diffusing out from the optically thick gas (hereafter ``cocoon cooling emission'') and non-thermal emission caused by the relativistic particles accelerated by the cocoon deceleration shocks (hereafter ``cocoon afterglow'', similar to jet afterglow). Due to differences in velocities, opening angles of different components, and durations of their emissions, observers from different lines of sight may observe a diversity of emission phenomena.

\subsection{Cocoon Dynamics and Cooling Emission} \label{sec:PhysicsOfCocoons}

We provide a unified treatment for each of the two cocoon components as an isotropic-equivalent expanding shell-like structure, with mass $M^{\rm iso}_{\rm c}$, kinetic energy $E_{\rm c,k}^{\rm iso}=(\Gamma_{\rm c}-1)M^{\rm iso}_{\rm c}c^2$, { initial volume at cocoon breakout} $V_{\rm c0}^{\rm iso}$, expansion speed $\beta_{\rm c}$, and the corresponding Lorentz factor $\Gamma_{\rm c}=(1-\beta_{\rm c}^2)^{-1/2}$. Hereafter, superscript ``${\rm iso}$'' means isotropic-equivalent --- the true mass $M_{\rm c}$ and kinetic energy $E_{\rm c,k}$ are smaller than the isotropic-equivalent values by a beaming factor $f_{\rm c} = \Omega_{\rm c}/(4\pi)$, where $\Omega_{\rm c}$ is the solid angle spanned by a given cocoon component. For the outer cocoon that is quasi-isotropic, we adopt a beaming factor of $f_{\rm oc}\approx 1$, whereas the beaming factor for the inner cocoon may be much smaller $f_{\rm ic}\ll 1$ due to relativistic motion.

{ Previous studies have developed models for the cooling emission \citep[e.g.][]{Nakar2017ApJ...834...28N,Gottlieb2022FBOT}, which are based on the energy distribution with velocity and the angle from the jet axis. In our approach, we simplify the picture by considering two main cocoon components (instead of a continuous distribution in velocities) based on their very different emissions in the X-ray and optical bands. We use isotropic equivalent quantities as those are directly constrained by observations. If our broad-brush model is confirmed by future observations, then this motivates further studies to construct a more detailed model with the full angular profile of the jet-cocoon system. 
} 

The two cocoon components are initially hot, radiation-dominated, and optically thick. At observer's time $t$ since beginning of the explosion, the radius of the cocoon is given by $R_{\rm c}\approx\beta_{\rm c}ct/(1-\beta_{\rm c})$. The corresponding optical depth is $\tau_{\rm c}\approx\kappa_{\rm R} M^{\rm iso}_{\rm c}/(4\pi R_{\rm c}^2)$, where $\kappa_{\rm R}$ is the Rosseland-mean opacity. For the inner cocoon where the gas is fully ionized, we take $\kappa_{\rm R}\approx 0.2\,{\rm cm^{2}g^{-1}}$, whereas for the outer cocoon which has a much lower temperature (see below), we take $\kappa_{\rm R}\approx 0.1\rm\, cm^2\,g^{-1}$. Our results are only weakly affected by these choices.

{ In the comoving frame, the thermal energy in the comoving frame decreases as $E'^{\rm iso}_{\rm c,th}\propto (V'^{\rm iso}_{\rm c})^{-1/3}$ due to adiabatic expansion.}
When the optical depth drops to $\tau_{\rm c}\approx \beta_{\rm c}^{-1}$, the diffusion timescale is comparable to the dynamical time and this occurs at a radius $R_{\rm c,diff}=\sqrt{\beta_{\rm c}\kappa_{\rm R} E^{\rm iso}_{\rm c,k}/4\pi(\Gamma_{\rm c}-1)c^2}$.
Thus, one can derive the photon diffusion time (i.e., the peak time) of the inner and outer cocoons
\begin{equation}
    \label{tdiff}
    \begin{split}
        t_{\rm c,diff} &= \sqrt{\frac{{\left(1 - \beta_{\rm c}\right)^2 \kappa_{\rm R} E^{\rm iso}_{\rm c,k}}}{{4\pi \beta_{\rm c}\left(\Gamma_{\rm c} - 1\right)c^4}}}  \\
    &\approx
    \begin{cases} 
        (24\,{\rm s})E_{\rm ic,k,51}^{1/2}f_{\rm ic,-1}^{-1/2}\Gamma_{\rm ic,1}^{-5/2}, &{\rm Inner}, \\ 
        (1.5\,{\rm d})E_{\rm oc,k,51}^{1/2}\beta_{\rm oc,-1}^{-3/2}, &{\rm Outer},
    \end{cases}
    \end{split}
\end{equation}
where $f_{\rm ic}=E_{\rm ic,k}/E_{\rm ic,k}^{\rm iso}$ is the beaming factor and $E_{\rm ic,k}$ is the beaming-corrected kinetic energy for the inner cocoon. In the second line of the above equation (and hereafter), we use approximations $\Gamma_{\rm ic}\gg 1$ for the inner cocoon and $\beta_{\rm oc}\ll 1$ for the outer cocoon.

{  The cocoon is initially non-relativistic when it breaks out from the progenitor star. Thus, its initial volume is the same in all Lorentz frames $V^{\rm iso}_{\rm c0}=V'^{\rm iso}_{\rm c0}$. We define { the volume-equivalent initial radius as}
\begin{equation}
    R_{\rm c0}\equiv(3V^{\rm iso}_{\rm c0}/4\pi)^{1/3}.
\end{equation}
{ Note that $R_{\rm c0}$ is comparable to the progenitor star's radius $R_{\star}$ --- the difference comes from the possible lateral expansion of the cocoon components after the breakout. This will be discussed in \S \ref{sec:progenitor} where we show $R_{\rm ic0}\approx R_{\star}$ for the inner cocoon and $R_{\rm oc0}\sim R_{\star}/3$  for the outer cocoon. }

Initially the cocoon is dominated by its thermal energy, but adiabatic expansion will then quickly convert nearly all the thermal energy into kinetic energy, so the initial thermal energy is given by $E^{\rm iso}_{\rm c,th}(R_{\rm c0}) \approx E^{\rm iso}_{\rm c,k}$, and the energy in the comoving frame is $ E'^{\rm iso}_{\rm c,th}(R_{\rm c0}) \approx E^{\rm iso}_{\rm c,th}(R_{\rm c0})$ as the cocoon is initially non-relativistic. 
In the observer's frame, the remaining thermal energy at the diffusion radius is $E^{\rm iso}_{\rm c,th}(R_{\rm c,diff})\approx \Gamma_{\rm c}E'^{\rm iso}_{\rm c,th}(R_{\rm c0})(V'^{\rm iso}_{\rm c0}/V'^{\rm iso}_{\rm c,diff})^{1/3}$, where the $\Gamma_{\rm c}$ comes from the Lorentz transformation from the comoving frame to the observer's frame. We adopt the volume at $R_{\rm c,diff}$ of the inner and outer cocoon as $V'^{\rm iso}_{\rm ic,diff}=2\pi R^3_{\rm ic,diff}/\Gamma_{\rm ic}$ and $V'^{\rm iso}_{\rm oc,diff}=4\pi R^3_{\rm ic,diff}/3$, respectively. } 
When the cocoon reaches the diffusion radius, the remaining thermal energy will be radiated away on a timescale of $t_{\rm c,diff}$, so we obtain the peak luminosity 
\begin{equation}
    \label{lumi}
    \begin{split}
    &L_{\rm c,peak} \approx E^{\rm iso}_{\rm c,th}(R_{\rm c,diff})/t_{\rm c,diff} \\
    &\approx
    \begin{cases} 
        5\times10^{48}\,R_{\rm ic0,11}\Gamma_{\rm ic,1}^{13/3}\,{\rm erg\,s^{-1}},&{\rm Inner}, \\
        2\times10^{42}\,R_{\rm oc0,11}\beta^{2}_{\rm oc,-1}\,{\rm erg\,s^{-1}},&{\rm Outer},
    \end{cases}
    \end{split}
\end{equation}
and the corresponding effective temperature (see Appendix \ref{appendix} for details)
\begin{equation}
\label{equ:temperature}
\begin{split}
    &T_{\rm c,eff} = \left[\frac{\Gamma^2_{\rm c}L_{\rm c,peak}}{8\pi R_{\rm c,diff}^2\sigma_{\rm SB}(1+\beta_{\rm c})}\right]^{1/4} \\
    &\approx\begin{cases} 
        (0.2\,{\rm keV}) E_{\rm ic,k,51}^{-1/4}f_{\rm ic,-1}^{1/4}
        R_{\rm ic0,11}^{1/4} 
        \Gamma^{11/6}_{\rm ic,1},&{\rm Inner}, \\
        (2 \times 10^4\,{\rm K}) E_{\rm oc,k,51}^{-1/4} 
        R_{\rm oc0,11}^{1/4}\beta_{\rm oc,-1}^{3/4},&{\rm Outer},
    \end{cases}
\end{split}
\end{equation}
where $\sigma_{\rm SB}$ is the Stefan-Boltzmann constant.

We find that the inner cocoon produces thermal cooling emission with a luminosity of the order $10^{48}\,R_{\rm ic0,11}\Gamma_{\rm ic,1}^{13/3}{\rm erg}\,{\rm s}^{-1}$, lasting for $10$ to $10^2$ seconds. The expected emission falls in the soft X-ray band (but see Section \ref{sec:Phase1} for potential correction to the observed color temperature by Comptonization). Such cooling emission can easily be detected by EP if the observer is along the relativistic beaming cone of the inner cocoon but outside the beaming cone of the GRB jet, i.e., the viewing angle relative to the jet axis is $\theta_{\rm j}\lesssim\theta_{\rm v}\lesssim\theta_{\rm ic}+\Gamma^{-1}_{\rm ic}$. The nearly isotropic cooling emission of the outer cocoon has a peak luminosity of the order $10^{42}\,R_{\rm oc0,11}\beta^{2}_{\rm oc,-1}\,{\rm erg\,s^{-1}}$ and duration of $\sim1\,{\rm day}$, making its brightness comparable to that of a typical SN but with a much faster evolution.  Note that the peak luminosity is strongly affected by the initial cocoon radius $R_{\rm c0}$ and the expansion speed. For instance, a much larger progenitor's radius $R_{\rm c0}\sim 100R_\odot$ than our fiducial value \citep[as considered by][in the context of low-luminosity GRBs]{nakar15_llGRB_envelope}, the peak luminosity may be as high as $10^{44}\rm\, erg\,s^{-1}$.
Additionally, the effective temperature of a few$\times10^4\,{\rm K}$ gives the optical emission a relatively blue color. After the peak time, both cocoon cooling emissions are expected to decay rapidly as $L_{\rm c}(t)\propto e^{-(t/t_{\rm c,diff})^2/2}$ \citep[e.g.,][]{Piro2021}.

\subsection{Afterglows}
The afterglow emission is produced by the (external) shocks formed between the jet/cocoon and the ambient medium. Initially, the bulk Lorentz factor of the shock-heated region remains constant until the deceleration time $t_{\rm dec}$ (defined in the observer's frame) when a large fraction of the initial kinetic energy is converted into thermal energy in the shock-heated region. The afterglow emission observed within the beaming cone before 
$t_{\rm dec}$ 
shows a rising ligthcurve for the interstellar medium (ISM) with a nearly constant hydrogen number density $n_{\rm ISM}$ and 
a shallow decay for the wind-like medium with number density following the conventional parametrization of $n(R) = 3\times10^{35} {\rm cm^{-1}}\, A_{\star}R^{-2}$ \citep{Zhang2018pgrb.book.....Z}.


\begin{figure*}
    \centering
    \includegraphics[scale=0.8]{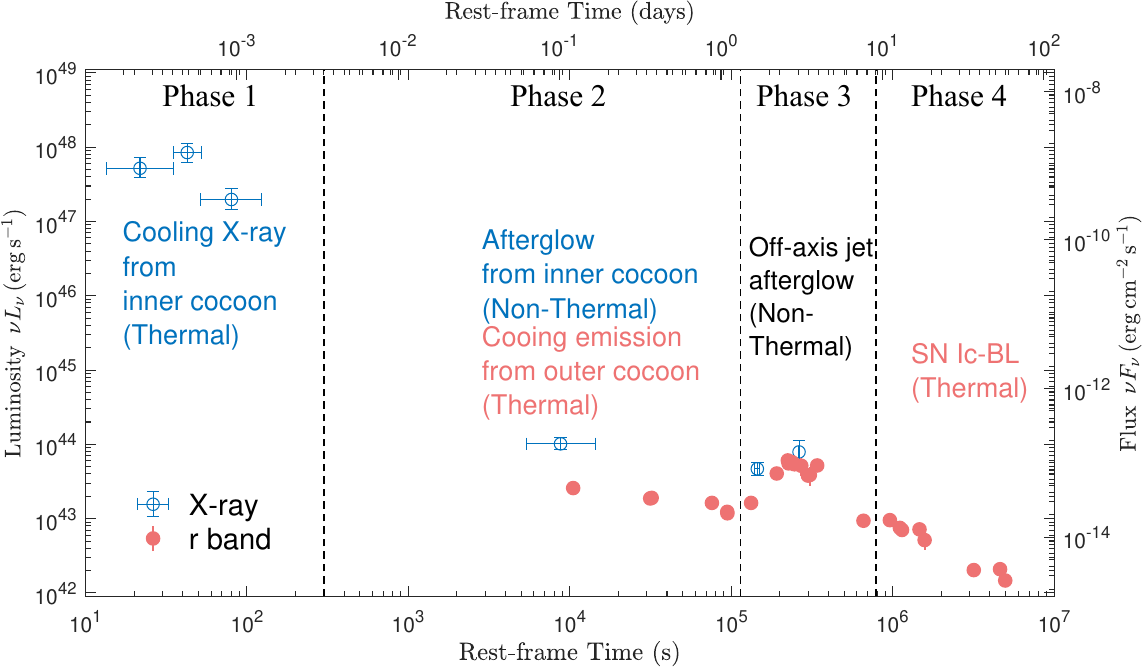}
    \caption{
    Four observed phases of EP240414a. The blue points represent the X-ray data from EP-WXT ($0.5-4\,$keV), EP-FXT ($0.5-10\,$keV) and Swift-XRT ($0.3-10\,$keV) \citep{Sun2024arXiv241002315S}, while the red points denote $r$ band data \citep{Srivastav2024arXiv240919070S,vanDalen2024arXiv240919056V}. The black dashed lines roughly indicate the boundaries between each phase. The proposed physical origins of the four phases are labeled in the figure.  
    }
    \label{fig:phase}
\end{figure*}

For an ultra-relativistic jet with initial Lorentz factor $\Gamma_{\rm j0} = 300\,\Gamma_{\rm j0,2.5}$, the deceleration time for an on-axis observer is $t_{\rm j,dec}\approx(26 \,{\rm s})E_{\rm j,k,51}^{1/3}f_{\rm j,-3}^{-1/3}n_{\rm ISM,0}^{-1/3}\Gamma_{\rm j0,2.5}^{-8/3}$ for the ISM case and $\approx(6\,{\rm s})E_{\rm j,k,51}f^{-1}_{\rm j,-3}A_{\star,-1}^{-1}\Gamma_{\rm j0,2.5}^{-4}$ for the wind case, where $f_{\rm j}$ is the jet beaming factor.
Thus, the deceleration time for an ultra-relativistic jet is typically less than 100 s, during which the jet afterglow may be very bright 
and likely outshine the cocoon emissions. The inner cocoon has a deceleration time of $t_{\rm ic,dec}\approx(0.6\,\mathrm{d})E_{\rm ic,k,51}^{1/3}f_{\rm ic,-1}^{-1/3}n_{\rm ISM,0}^{-1/3}\Gamma_{\rm ic0,1}^{-8/3}$ in the ISM and $\approx(0.5\,\mathrm{d})E_{\rm ic,k,51}f^{-1}_{\rm ic,-1}A_{\star,-1}^{-1}\Gamma_{\rm ic0,1}^{-4}$ in the wind medium, and the cocoon afterglow can only be observed if the jet afterglow is weak enough at this time.

After the deceleration time, the shock evolution transitions into the self-similar phase during which the afterglow as observed by an on-axis observer begins to fade in most bands above the minimum characteristic synchrotron frequency  or $\nu>\nu_{\rm m}$. If the beaming angle is still narrower than the opening angle of the jet or cocoon (i.e. $\Gamma^{-1}_{\rm \{j,c\}}<\theta_{\rm \{j,c\}}$), the flux decays as a power-law $t^{\alpha}$ with $\alpha=-3(p-1)/4$ in the ISM case and $-(3p-1)/4$ in the wind case \citep{Gao2013NewAR..57..141G,Zhang2018pgrb.book.....Z}, where $p$ is the power-law index of the electron Lorentz factor distribution.
There are two jet effects: (1) the edge effect \citep[e.g.][]{Panaitescu1998ApJ...503..314P}, which steepens lightcurves by an additional factor of $t^{-\frac{3}{4}}$ in the ISM case ($t^{-\frac{1}{2}}$ in the wind case), and (2) the lateral expansion of the jet or cocoon, causing a ``jet break'' and the flux to decay as $t^{-p}$ \citep{Rhoads1999ApJ...525..737R,Sari1999ApJ...519L..17S}.
An on-axis observer can detect the jet afterglow's evolution from the coasting phase to self-similar phase at early time, with cocoon afterglow potentially emerging during the jet break phase.

For a slightly off-axis observer (i.e., $\theta_{\rm j}<\theta_{\rm v}\lesssim\theta_{\rm ic}+\Gamma^{-1}_{\rm ic}$), the jet afterglow is initially unobservable \citep{Beniamini2020MNRAS.492.2847B}. The inner cocoon afterglow may dominate the early-time non-thermal emission, displaying the coasting and then self-similar phases. Meanwhile, the jet afterglow rapidly rises as it decelerates, peaks when the beaming angle is comparable to the viewing angle $\theta_{\rm v}$, and then evolves similarly as an on-axis post-jet-break afterglow.

\section{Application to EP240414a}
\label{sec:app}

Hereafter, all observed parameters, which we list in Table \ref{tab:parameters}, are represented in the host galaxy's rest frame. Based on the description in the Introduction, one can divide the evolution of EP240414a's optical and X-ray emissions into four distinct phases: (see Figure \ref{fig:phase})
\begin{itemize}
  \item{Phase 1}: prompt X-ray emission for $\sim\!110\,{\rm s}$;
  \item{Phase 2}: non-thermal X-ray emission around $0.1\,{\rm d}$ transitioning into the thermal-dominated optical plateau lasting for $\sim\!1$\,d;
  \item{Phase 3}: non-thermal optical bump peaking at $3$\,d;
  \item{Phase 4}: SN Ic-BL emerging after 7\,d.
\end{itemize}

We first consider the origin of the prompt X-ray emission in Phase 1. The very soft spectra (average photon index $-3.1^{+0.7}_{-0.8}$), low peak energy $E_{\rm p}<1.3$\,keV, and the absence of strong temporal variability all differ from the prompt emission from typical GRB jets \citep{Sun2024arXiv241002315S}.

The blackbody model provides a comparably good fit as the power-law model \citep{Sun2024arXiv241002315S}, supporting a thermal origin. One possibility is the shock breakout. However, we find that the peak luminosity $1.3\times10^{48}\,{\rm erg\,s^{-1}}$, peaking at  $\sim\!70\,{\rm s}$, and temperature $\sim\!0.5\,{\rm keV}$ \citep{Sun2024arXiv241002315S} deviate significantly from the so-called breakout closure relation suggested by \cite{Nakar2012ApJ...747...88N}. Alternatively, these properties may be explained by: (1) an on-axis weak jet with a lower Lorentz factor $\Gamma_{\rm j}\ll 100$
which does not produce bright GRB prompt emission; 
(2) the inner cocoon emission. In the on-axis weak jet scenario, the $\sim\!0.1\,{\rm d}$ (X-ray) and $\sim\! 3\,{\rm d}$ (optical) non-thermal components may be attributed to the jet afterglow and cocoon afterglow, respectively. However, { spectroscopic observations indicate the optical flux after 7 days is likely dominated by the SN Ic-BL \citep{vanDalen2024arXiv240919056V}, suggesting a rapid decay of the $\sim\!3\,{\rm d}$ optical bump.}
We find that decaying slope is inconsistent with the on-axis cocoon afterglow; instead, it is more consistent with the afterglow from an off-axis GRB jet (see Section \ref{sec:Phase3}).

Furthermore, the observed peak luminosity, duration, and spectrum of the prompt X-ray emission in EP240414a are broadly consistent with our predictions for the inner cocoon cooling emission in Section \ref{sec:PhysicsOfCocoons}. As for the early thermal-dominated optical emission, we find that the luminosity of $\sim\!2\times10^{43}\,{\rm erg}\,{\rm s}^{-1}$, diffusion time of $\sim\!0.4\,{\rm d}$, and the fitted blackbody temperature of $24000\pm7000\,{\rm K}$ \citep{vanDalen2024arXiv240919056V} in Phase 2 may be reasonably explained by the outer cocoon cooling emission as shown in Section \ref{sec:PhysicsOfCocoons}. 

In the following, we provide a more detailed analysis of the origins of the different components in each phase and constrain the properties of the jet, progenitor, and nearby environment based on observations.

\begin{deluxetable*}{c|ccc|ccc}[tb]
\tabletypesize{\footnotesize}
\tablecaption{Rest-frame Observables and Constrained Properties of Cocoons} 
\label{tab:parameters}
\tablehead{
\colhead{Cocoon} &
\colhead{$L_{\rm c,peak}({\rm erg\,s^{-1}})$} &
\colhead{$t_{\rm c,diff}$} &
\colhead{$T_{\rm c,peak} $} &
\colhead{$E^{\rm iso}_{\rm c,k}$(erg)} &
\colhead{$\Gamma_{\rm ic0}/\beta_{\rm oc}$} &
\colhead{$R_{\rm c0}(R_{\odot})$} 
}
\startdata
    Inner & $1.3^{+0.4}_{-0.4}\times10^{48}$ & 70\,s & $0.5^{+0.06}_{-0.06}\, {\rm keV}$ & $1.0^{+0.1}_{-0.1}\times10^{52}$ & $\Gamma_{\rm ic0}=6.5^{+0.1}_{-0.1}$ & $2.5^{+0.2}_{-0.2}$\\
    Outer ($f_{\tau}=1.0$) & $1.9^{+0.1}_{-0.1}\times10^{43}$ & 0.4\,d & $3.4^{+1.0}_{-1.0}\times10^{4}\,{\rm K}$  & $0.4^{+4.7}_{-0.3}\times10^{50}$ & $\beta_{\rm oc}=0.08^{+0.09}_{-0.03}$ & $22.3^{+24.5}_{-17.8}$\\
    Outer ($f_{\tau}=1.5$) & - & - & -  & $4.4^{+35.4}_{-3.0}\times10^{50}$ & $\beta_{\rm oc}=0.16^{+0.14}_{-0.05}$ & $5.0^{+5.5}_{-3.8}$\\
\enddata
\tablecomments{The observed values of the parameters for the two cocoons are taken from \cite{Sun2024arXiv241002315S} and \cite{vanDalen2024arXiv240919056V}. The diffusion time $t_{\rm c,diff}$ of the two cocoon components are taken as the peak time of the lightcurves, and their  uncertainties are treated as systematic and not included in our statistical inference.  }

\end{deluxetable*}

\subsection{Phase 1: Inner Cocoon Cooling Emission (X-ray)} \label{sec:Phase1}

Following the above arguments, the X-ray prompt emission of EP240414a could originate from the thermal cooling emission of the mildly relativistic inner cocoon. Based on the measured peak bolometric luminosity $L_{\rm ic,peak}$, peak color temperature $T_{\rm ic,peak}$ (which is generally different from but related to the effective temperature $T_{\rm ic,eff}$ in Eq. \ref{equ:temperature}), and duration $t_{\rm ic,diff}$, one may use Equations (\ref{tdiff}), (\ref{lumi}), and (\ref{equ:temperature}) to constrain the isotropic kinetic energy $E_{\rm ic,k}^{\rm iso}$, initial Lorentz factor $\Gamma_{\rm ic0}$, and the initial radius $R_{\rm ic0}$.

However, dilute, high temperature plasma may deviated from local thermodynamic equilibrium (LTE) if insufficient photons are emitted to thermalize the radiation field. Instead, photons and electrons will reach Compton equilibrium with an observed color temperature at peak luminosity $T_{\rm ic,peak}$ higher than the effective temperature $T_{\rm ic,eff}$.

As bremsstrahlung is the dominating emission process, most photons are produced near the beginning of the expansion at $V'^{\rm iso}_{\rm ic0}$. Thus, we compare the radiation energy density $a{T^{\prime\,4}_{\rm ic0}}$ with that produced by free-free emission $\epsilon^{\prime}_{\rm ff}t^{\prime}_{\rm dyn}$ in the comoving frame at { the initial volume $V'^{\rm iso}_{\rm ic0}$}, where $a$ is the radiation density constant, { $T'_{\rm ic0}=T_{\rm ic,eff}(V^{\rm iso}_{\rm ic,diff}/V'^{\rm iso}_{\rm ic0})^{1/3}\Gamma^{-1}_{\rm ic0}$} is the blackbody-equivalent radiation temperature, $\epsilon^{\prime}_{\rm ff}\approx1.4\times10^{-27}\sqrt{T^{\prime}_{\rm ic0}}{n_0^{\prime}}^2\,{\rm erg\,cm^{-3}\,s^{-1}}$ is the emissivity at temperature $T_{\rm ic0}'$, { $n^{\prime}_0=M^{\rm iso}_{\rm ic}/( V'^{\rm iso}_{\rm ic0}m_{\rm p})\propto E^{\rm iso}_{\rm ic,k}R^{-3}_{\rm c0}$ is the number density, and $t^{\prime}_{\rm dyn}\approx R_{\rm ic0}/c$ is the dynamical timescale right after the breakout (trans-relativistic).} Thus, one can define the ratio between the two energy densities
\begin{equation}
    \eta_0\equiv {aT_{\rm ic0}^{\prime\,4}\over \epsilon'_{\rm ff}t'_{\rm dyn}} \approx1.6\,E^{-9/8}_{\rm ic,k,51}f_{\rm ic,-1}^{9/8}R_{\rm ic0,11}^{19/8}\Gamma^2_{\rm ic0,1},
\end{equation}
which is highly sensitive to $R_{\rm ic0}$. If the ratio $\eta_0>1$, free-free emission is insufficient to thermalize the radiation field, leading to a photon-starved Compton equilibrium with a fixed photon number along the expansion of the inner cocoon. 
The observed color temperature is given by \citep{Nakar2010ApJ...725..904N}
\begin{equation}
\label{equ:ModifiedTemperature}
    T_{\rm ic,peak}=T_{\rm ic,eff}\max(\eta^2_0,1).
\end{equation}
Thus, $L_{\rm ic,peak}$, $T_{\rm ic,peak}$, and $t_{\rm ic,diff}$,along with Equations (\ref{tdiff}), (\ref{lumi}), (\ref{equ:temperature}) and (\ref{equ:ModifiedTemperature}), allow us to solve $E_{\rm ic,k}^{\rm iso}$, $\Gamma_{\rm ic0}$, and $R_{\rm ic0}$ for the inner cocoon. We find 
\begin{equation}
\eta_0\approx2.5\,\left(\frac{{L}_{\rm ic,peak, 48}}{1.3}\right)^{-\frac{61}{728}}\left(\frac{{t}_{\rm ic,diff}}{70\,{\rm s}}\right)^{\frac{19}{91}}\left(\frac{{T}_{\rm ic,peak}}{0.5\,{\rm keV}}\right)^{\frac{179}{364}}, 
\end{equation}
suggesting that the X-ray spectrum from the inner cocoon cooling emission is in the photon-starved regime. The inferred physical parameters of the inner cocoon are 
\begin{equation}
    \label{equ:Gamma}
    \begin{split}
    &\Gamma_{\rm ic0}\approx\eta^4_0\left(\frac{{L}_{\rm ic,peak}}{64\pi \sigma_{\rm SB}{T}^4_{\rm ic,peak}c^2{t}^2_{\rm ic,diff}}\right)^{1/2} \\
    &\approx 6.5\left(\frac{{L}_{\rm ic,peak, 48}}{1.3}\right)^{\frac{15}{91}}\left(\frac{{t}_{\rm ic,diff}}{70\,{\rm s}}\right)^{-\frac{15}{91}}\left(\frac{{T}_{\rm ic,peak}}{0.5\,{\rm keV}}\right)^{-\frac{3}{91}},
\end{split}
\end{equation}
\begin{equation}
\begin{split}
    &E^{\rm iso}_{\rm ic,k}\approx16\pi\kappa^{-1}_{\rm R}\Gamma^5_{\rm ic0}c^4{t}^2_{\rm ic,diff}\approx 10^{52} {\rm\, erg}\\
    &\left(\frac{{L}_{\rm ic,peak,48}}{1.3}\right)^{\frac{75}{91}} 
    \left(\frac{{t}_{\rm ic,diff}}{70\,{\rm s}}\right)^{\frac{107}{91}}\left(\frac{{T}_{\rm ic,peak}}{0.5\,{\rm keV}}\right)^{-\frac{15}{91}},
\end{split}
\end{equation}
and
\begin{equation}
\begin{split}
    &R_{\rm ic0}\approx\frac{2{L}_{\rm ic,peak}\Gamma^{2/3}_{\rm ic0}c\,{t}^2_{\rm ic,diff}}{0.8E^{\rm iso}_{\rm ic,k}}\\
    &\approx 2.5\, R_{\odot}\left(\frac{{L}_{\rm ic,peak,48}}{1.3}\right)^{\frac{2}{7}}
    \left(\frac{{t}_{\rm ic,diff}}{70\,{\rm s}}\right)^{\frac{5}{7}}\left(\frac{{T}_{\rm ic,peak}}{0.5\,{\rm keV}}\right)^{\frac{1}{7}}.
\end{split}
\end{equation}
With these physical parameters, the predicted peak photon energy is

\begin{equation}
\begin{split}
    E_{\rm p}&\approx 3k{T}_{\rm ic,peak} \\
    &\approx1.5 {\rm\, keV}\, E^{\rm iso\,-5/2}_{\rm ic,k,52}
    \left(\frac{{R}_{\rm ic0}}{2.5\,R_{\odot}}\right)^{5}\left(\frac{{\Gamma}_{\rm ic0}}{6.5}\right)^{\frac{35}{6}},
\end{split}
\end{equation}
and isotropic X-ray energy is
\begin{equation}
\begin{split}
    E^{\rm iso}_{\rm X}&\approx E^{\rm iso}_{\rm ic,k}(0.8\Gamma^{4/3}_{\rm ic0} R_{\rm ic0}/R_{\rm ic,diff})\\
    &\approx10^{50}{\rm \,erg}\, E^{\rm iso\,1/2}_{\rm ic,k,52}
    \left(\frac{{R}_{\rm ic0}}{2.5\,R_{\odot}}\right)\left(\frac{{\Gamma}_{\rm ic0}}{6.5}\right)^{\frac{11}{6}}.
\end{split}
\end{equation}
Both of them are consistent with observations. 

Taking into account the uncertainties of the observables, the inferred median values for the parameters along with their $1\sigma$ credible intervals are shown in Table \ref{tab:parameters}. The isotropic cocoon energy $E^{\rm iso}_{\rm ic,k}=1.0^{+0.1}_{-0.1}\times10^{52}$\,erg, which corresponds to a beaming-corrected energy of $E_{\rm ic,k}\sim10^{51}\,{\rm erg}$ for a typical beaming factor of $f_{\rm ic}\sim0.1$. The energy of the inner cocoon is expected to be comparable to that of the GRB jet \citep[e.g.,][]{Nakar2017ApJ...834...28N}.
{ The initial radius is $R^{\rm}_{\rm ic0}=2.5^{+0.2}_{-0.2}R_{\odot}$, supporting the progenitor of EP240414a as a stellar object.}
The isotropic mass of the inner cocoon is $M_{\rm ic}^{\rm iso}=1.0^{+0.1}_{-0.1}\times10^{-3}M_{\odot}$ and the beaming corrected mass is an order of magnitude smaller. 


\subsection{Phase 2: Outer Cocoon Cooling Emission (optical)}\label{sec:phase2_outer_cocoon}

Since the outer cocoon is much denser (than the inner cocoon), LTE holds in its interior ($\eta_0\approx10^{-6}E^{-9/8}_{\rm oc,k,51}R^{19/8}_{\rm oc0,11}\beta^{5}_{\rm oc0,-1}$). We ignore the difference between the thermalization radius and the photosphere, and take the observed { color temperature near peak luminosity to be $T_{\rm oc,peak} \approx f_{\tau} T_{\rm oc,eff}$, where $1<f_{\tau}\lesssim1.5$ is the color correction factor for the outer cocoon (see Appendix \ref{appendix} for details).}

Based on the observationally inferred ${L}_{\rm oc,peak}$, ${T}_{\rm oc,peak}$, and ${t}_{\rm oc,diff}$ for the outer cocoon cooling emission, one can constrain the physical parameters through Equations (\ref{tdiff}), (\ref{lumi}), and (\ref{equ:temperature}) as follows
\begin{equation}
    \label{equ:beta}
    \begin{split}
    &\beta_{\rm oc0}\approx\left(\frac{{L}_{\rm oc,peak}}{8\pi \sigma_{\rm SB}{T}^4_{\rm oc,peak}c^2{t}^2_{\rm oc,diff}}\right)^{1/2}\approx 0.1 \\
    &\left(\frac{{L}_{\rm oc,peak,43}}{1.9}\right)^{\frac{1}{2}}\left(\frac{{t}_{\rm oc,diff}}{0.4\,{\rm d}}\right)^{-1}\left(\frac{{T}_{\rm oc,peak,4}/f_{\tau}}{3.4}\right)^{-2},
\end{split}
\end{equation}
\begin{equation}
\begin{split}
    &E^{\rm iso}_{\rm oc,k}\approx2\pi\kappa^{-1}_{\rm R}\beta^3_{\rm oc0}c^4{t}^2_{\rm oc,diff}\approx 5\times10^{49}{\rm\, erg} \\
    &\left(\frac{{L}_{\rm oc,peak,43}}{1.9}\right)^{\frac{3}{2}}\left(\frac{{t}_{\rm oc,diff}}{0.4\,{\rm d}}\right)^{-1}\left(\frac{{T}_{\rm oc,peak,4}/f_{\tau}}{3.4}\right)^{-6},
\end{split}
\end{equation}
and
\begin{equation}
\begin{split}
    R_{\rm oc0}&\approx\frac{{L}_{\rm oc,peak}\beta_{\rm oc0}c\,{t}^2_{\rm oc,diff}}{E^{\rm iso}_{\rm oc,k}}\\
    &\approx20\,R_{\odot}\left(\frac{{t}_{\rm oc,diff}}{0.4\,{\rm d}}\right)^{2}\left(\frac{{T}_{\rm oc,peak,4}/f_{\tau}}{3.4}\right)^{4},
\end{split}
\end{equation}
Including the $1\sigma$ uncertainties of the observables, the outer cocoon { without color correction ($f_{\tau}=1$)} has kinetic energy $E_{\rm oc,k}\approx0.4^{+4.7}_{-0.3}\times10^{50}{\rm erg}$, dimensionless velocity of $\beta_{\rm oc}\approx0.08^{+0.09}_{-0.03}$, and initial radius $R_{\rm oc0}=22.3^{+24.5}_{-17.8}\,R_\odot$, as also listed in Table \ref{tab:parameters}. One can thus infer the outer cocoon mass to be $M_{\rm oc}= 7.6^{+11.8}_{-2.7}\times10^{-3}M_{\odot}$. { Adopting the maximum color correction of $f_{\tau}=1.5$, the inferred physical parameters are $E_{\rm oc,k}\approx4.4^{+35.4}_{-3.0}\times10^{50}{\rm erg}$, $\beta_{\rm oc}\approx0.16^{+0.14}_{-0.05}$, $R_{\rm oc0}=5.0^{+5.5}_{-3.8}\,R_\odot$, and $M_{\rm oc}= 18.2^{+28.7}_{-6.8}\times10^{-3}M_{\odot}$.}

We note that these constraints are very sensitive to the observationally inferred temperature as well as the model assumptions on the predicted color temperature.
Contamination from afterglow emission in the near-infrared bands may cause the temperature inferred by \cite{vanDalen2024arXiv240919056V} based on a blackbody fit to be underestimated. On the other hand, if the thermalization radius is much smaller than the photosphere, the model-predicted color temperature would be higher than the effective temperature { leading to $f_{\tau}>1$}. Moreover, the diffusion time of the outer cocoon $t_{\rm oc,diff}$ may have an uncertainty up to a factor of $\sim$2 mainly due to insufficient observing cadence. { Finally, the unknown host extinction also affects the inferred temperature of the outer cocoon's cooling emission.} For these reasons, our constraints on the outer cocoon parameters may have large systematic errors that are not captured by the (already very large) statistical errors in Table \ref{tab:parameters}.

\subsection{Afterglows from Inner Cocoon and Off-axis Jet} 
\label{sec:Phase3}

The featureless and flat optical spectrum observed near $3$\,d suggests a non-thermal origin \citep{Srivastav2024arXiv240919070S,vanDalen2024arXiv240919056V}. Hereafter, we adopt the $F_{\nu}\propto\nu^{\beta}t^{\alpha}$ notation for the spectral index $\beta$ and temporal power-law index $\alpha$.

First, we argue that the rapid evolution of the optical bump, with a combination of a rising slope of $\alpha = 2.4$ to  4 between $\sim\!1-3\,{\rm d}$ and a decaying slope from $\alpha = -2.2$ to $-2.6$ between $\sim\!3-7\,{\rm d}$, is inconsistent with the expected behavior of ordinary on-axis afterglows. 
The rising slope might be reproduced by an on-axis jet propagating in the coasting phase (before $t_{\rm dec}$) in the ISM environment with $\alpha=3$, while the rapid decay favors a jet in the lateral expansion phase with $\alpha=-p$. However, fine-tuning is required for the jet break to follow immediately after the deceleration. Another plausible explanation is the reverse shock, which can produce a rapid rise before the reverse shock crosses the entire jet and then a steep decline afterwards \citep{Gao2013NewAR..57..141G}. However, it is unusual
for the reverse shock crossing to occur on such a long timescale \citep[but see][]{Abdikamalov2025_2025arXiv250212757A}. Generally, the reverse shock crossing time is of the order the deceleration time or shorter, and an unusually weak jet with $\Gamma_{\rm j0}\lesssim 10$ is required to match the observed peak time.
Furthermore, in this scenario, the slower-evolving forward shock emission, which should emerge after the reverse shock crossing, was not observed. Thus, we conclude that it is challenging to explain the optical bump with the afterglow from an on-axis jet.

On the other hand, an off-axis jet afterglow can provide a self-consistent explanation for the rapid-evolving optical bump.  
An off-axis observer may detect the afterglow in the jet-break phase prior to its peak, characterized by a rising slope of $\alpha\approx4$ and a decaying slope of $\alpha=-3p/4$ (or $\alpha=-p$) due to the jet edge effect (or lateral expansion expansion). These are consistent with observations.
The off-axis afterglow emission reaches its peak when the jet Lorentz factor has decelerated to $\Gamma_{\rm j}(t)\approx\theta^{-1}_{\rm v}$. In the self-similar phase, the Lorentz factor decreases as $\Gamma_{\rm j} (t)\propto t^{-3/8}$ for the ISM environment, and steepens to $\Gamma_{\rm j}(t)\propto t^{-1/2}$ during the lateral expansion phase (after $\Gamma_{\rm j}(t)\approx\theta^{-1}_{\rm j0}$). Thus, $\theta_{\rm v}$ can be roughly inferred as
\begin{equation}
\label{equ:Theta_v}
    \theta_{\rm v}\approx15^{\circ}E^{-1/6}_{\rm j,k,51}n^{1/6}_{\rm ISM,0}({t}_{\rm peak}/3\,{\rm d})^{1/2}, 
\end{equation}
where ${t}_{\rm peak}$ is the observed peak time of the bump.  

\begin{figure}
    \centering
    \includegraphics[width=1.0\linewidth]{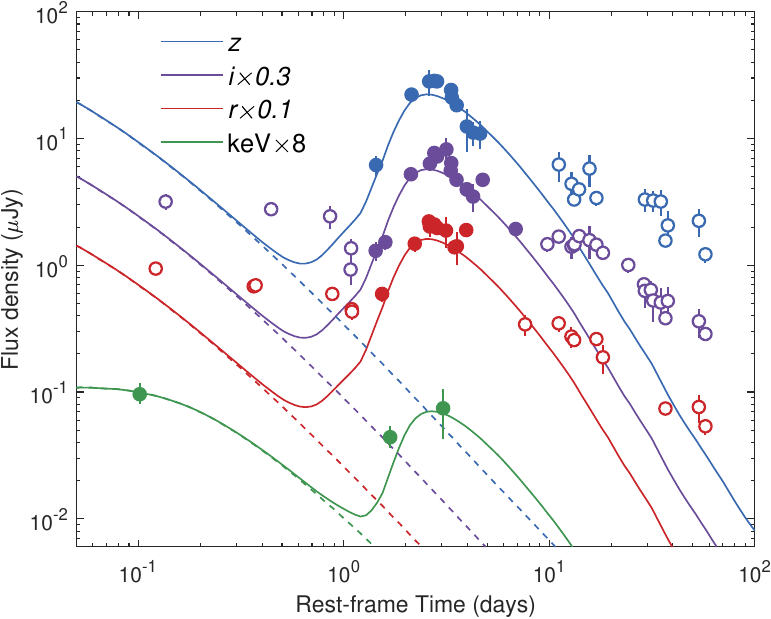}
    \includegraphics[width=1.0\linewidth]{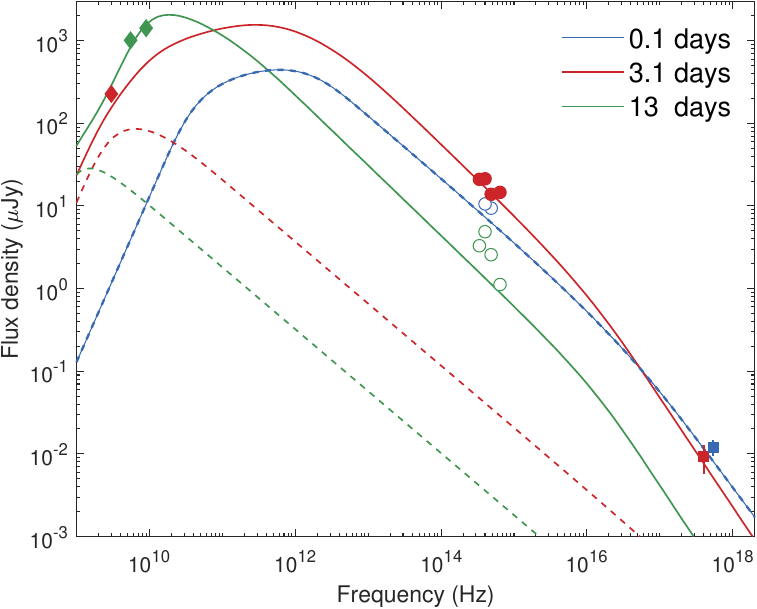}
    \caption{Multi-wavelength fitting for EP240414a. The upper panel shows the lightcurves in $z$ band (blue), $i$ band (purple), $r$ band (red) and 1\,keV X-rays (green). The lower panel displays the spectral energy distributions at $0.1\,{\rm days}$ (blue line), $4.3\,{\rm days}$ (red line), and $18.5\,{\rm days}$ (green line). { Filled symbols are afterglow-dominated data points and open symbols mark the fluxes dominated by cooling emissions and SN Ic-BL.} Data are collected from \cite{Sun2024arXiv241002315S}, \cite{Srivastav2024arXiv240919070S}, and \cite{vanDalen2024arXiv240919056V}, with Galactic extinction corrected.
    The solid lines show the sum of the afterglow contributions from the jet and inner cocoon, while the dashed lines are for the inner cocoon afterglow alone. In the lower panel, the blue dashed line at 0.1\,d overlaps with the solid line, as the inner cocoon afterglow dominates.
    }
    \label{afterglow}
\end{figure}

We propose a simple model in which both the jet and inner cocoon have a top-hat structure. Assuming a broken power-law density profile that transitions from a stellar wind  at small radii to a uniform ISM environment, we fit the X-ray and optical data between $1$ and $7\,{\rm days}$, as well as the late-time radio data, using our afterglow model (see details in Appendix \ref{appendix2}). We also fit the non-thermal dominated X-ray at $0.1\,$days with the inner cocoon afterglow, and the contemporary (thermal-dominated) optical fluxes are used as upper limits.  The lateral expansion of both the jet and inner cocoon has been included in our calculations.

\begin{figure}
    \centering
    \includegraphics[width=1.0\linewidth]{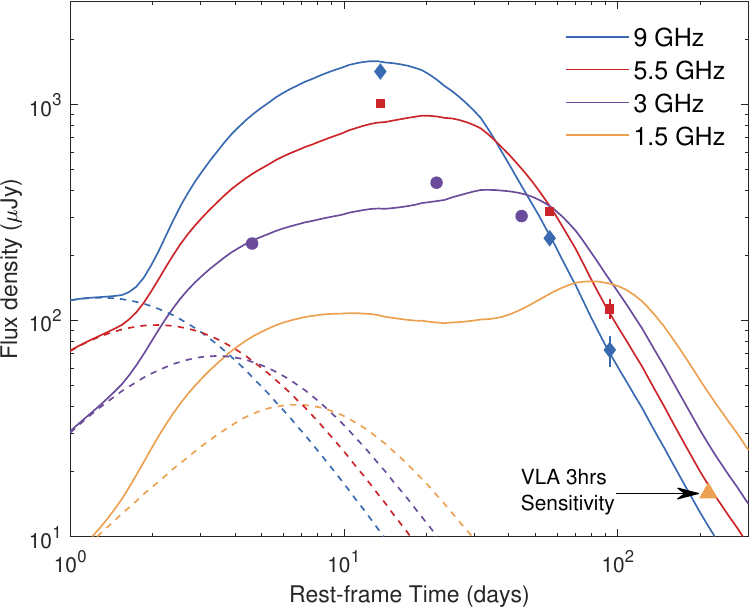}
    \caption{Radio afterglow lightcurves at 9\,GHz (blue diamond), 5.5\,GHz (red square) and 3\,GHz (purple circles). The model prediction at 1.5 GHz is shown by the orange line, together with the 3-hr VLA sensitivity as a yellow triangle \citep{Perley2011ApJ...739L...1P}. Data are taken from \cite{Bright2024arXiv240919055B} and \cite{Sun2024arXiv241002315S}. The dashed lines are for inner cocoon afterglow only and solid lines are for the total flux.}
    \label{afterglow2}
\end{figure}

The fits to the multi-wavelength lightcurves and spectra are shown in Figures \ref{afterglow} and \ref{afterglow2}. Afterglow emission models with a viewing angle between $\theta_{\rm v}=10^\circ-15^\circ$ can provide comparable fits to the available dataset, roughly consistent with our rough estimation in Equation (\ref{equ:Theta_v}).  The prompt gamma-ray emission from the top-hat jet would be too faint to be detected at viewing angles $\theta_{\rm v}\geq10^{\circ}$ if the jet Lorentz factor is $\Gamma_{\rm j}\geq30$ (see Appendix \ref{appendix3} for details).

We choose the median value $\theta_{\rm v}=12^{\circ}$ to show our results. The corresponding density profile parameters are $A_{\star}=0.1$ and $n_{\rm ISM}=0.6\,{\rm cm^{-3}}$, and the wind transitions to the ISM environment at a radius of $\sim\!2\times10^{17}\,{\rm cm}$. For the fitting result, the inner cocoon's parameters are: isotropic energy $E_{\rm ic,k}^{\rm iso}=10^{52}\,{\rm erg}$,  electron energy fraction $\epsilon_{e,\rm ic}=0.12$, magnetic energy fraction $\epsilon_{\rm B,ic}=1.5\times10^{-4}$, initial half opening angle $\theta_{\rm ic0}=20^\circ$, electron power-law index $p_{\rm ic}=2.5$, and initial Lorentz factor of $\Gamma_{\rm ic0}=6.5$. The isotropic energy and initial Lorentz factor of the inner cocoon are consistent with the values inferred from the prompt X-ray emission shown in Section \ref{sec:Phase1}. The values of these parameters for the off-axis jet are: $E^{\rm iso}_{\rm j,k}=7\times10^{53}$erg, $\epsilon_{e,\rm j}=0.12$, $\epsilon_{\rm B,j}=5\times10^{-3}$, $\theta_{\rm j0}=2.5^{\circ}$, $\Gamma_{\rm j0}=300$, and $p_{\rm j}=2.7$. We did not carry out a comprehensive exploration through the entire parameter space, as our purpose is to demonstrate the plausibility of our overall picture.

Within the wind environment, the jet break time for the inner cocoon is roughly $0.2\,{\rm d}$. Thus, the cocoon afterglow emission decays with a shallow slope before 0.2\,d, producing a plateau in the X-ray band and contributing nearly half of the $0.1\,$d optical flux.
The inner cocoon afterglow then transitions into the jet break phase, leading to a much faster decay.

On the other hand, the jet enters the ISM around $\sim\!0.1$\,d and generates bright ``orphan'' afterglow emission that dominates the $\sim\!3\,{\rm d}$ optical bump and the contemporary X-ray flux.
During the optical bump, the synchrotron cooling frequency $\nu_{\rm c}$ lies between the optical and X-ray bands. 
The model-predicted spectral indices are $\beta=-0.85$ in the optical and $\beta=-1.35$ in the X-ray, which are consistent with the observed optical-to-X-ray spectral slope of $\beta_{\rm ox}=-1.1\pm0.1$ \citep{vanDalen2024arXiv240919056V}.

The radio afterglow is shown in Figure \ref{afterglow2}. The inner cocoon afterglow dominates the radio emission before $\sim\!2\,{\rm d}$ when the radio emission from the jet is highly suppressed due to relativistic beaming. The rapid rise in the jet afterglow emission between 3 and 10\,d is caused by the combination of the widening beaming cone and the decreasing minimum frequency $\nu_{\rm m}$ and self-absorption frequency $\nu_{\rm a}$. The rapid flux drop after $\sim$30\,d is caused by the self-absorption frequency crossing the observing band, and the optically thin afterglow flux follows the same $t^{-p}$ evolution as the optical afterglow.

As $\nu_{\rm a}$ crosses the lowest frequency band the latest, we provide a model prediction for the 1.5 GHz lightcurve in Figure. \ref{afterglow2}, which shows that the source will remain above the 3-hr detectability limit of the Very Large Array (VLA) for about 1 year in the observer's frame.

\section{Summary and Discussion} \label{sec:Summary}
We propose that EP240414a, associated with an SN {\rm I}c-BL, {\rm likely} originates from a jet-cocoon system observed at a viewing angle $10^{\circ}\lesssim \theta_{\rm v}\lesssim 15^\circ$. The prompt X-ray emission arises from the thermal cooling of the inner cocoon with initial Lorentz factor $\Gamma_{\rm ic0}\approx 7$, isotropic kinetic energy $E^{\rm iso}_{\rm ic,k}\approx 10^{52}$\,erg, and initial radius $R_{\rm ic0}\approx3\,R_{\odot}$. 
The afterglow from the inner cocoon accounts for the non-thermal X-ray emission and about half of the optical flux at $\sim\!0.1\,{\rm d}$. The outer cocoon produces bright thermal emission 
that dominates the $\sim\!0.4\,$d optical plateau. We constrain the outer cocoon parameters as $\beta^{\rm}_{\rm oc}\sim 0.1f^{2}_{\tau}$, $10^{49}f^{6}_{\tau}\lesssim E_{\rm oc,k}\lesssim 5f^{6}_{\tau}\times 10^{50}$\,erg, and $4.5f^{-4}_{\tau}\lesssim R_{\rm oc0}\lesssim 47f^{-4}_{\tau}\,R_\odot$. We note that, due to systematic (both theoretical and observational) uncertainties of the color temperature of the outer cocoon cooling emission, these inferred parameters have large uncertainties, but they overall show that our model is plausible.

The optical bump peaking at 3\,days and subsequent bright radio emissions are attributed to the off-axis jet afterglow.
We constrain the beaming-corrected energy of the off-axis jet and the inner cocoon to be both of the order $E_{\rm j,k}\sim E_{\rm ic,k}\sim 10^{51}{\rm erg}$. This is consistent with studies that suggest the energies of the LGRB jet and cocoon should be comparable, typically on the order of $10^{51}$\,erg \citep[e.g.][]{Cenko2010ApJ...711..641C}.
According to our model, EP240414a joins the small sample of sources with well-observed off-axis jet afterglow emission, including XRF020903 \citep[e.g.,][]{Soderberg2004ApJ...606..994S,Lamb2005ApJ...620..355L,Urata2015ApJ...806..222U} and GRB170817A \citep[e.g.,][]{{Nathanail2020MNRAS.495.3780N,2021ARA&A..59..155M,mooley22_GW170817}}. { A newly detected source, EP241021a, also appears to exhibit off-axis jet afterglow emission \citep{Busmann2025arXiv250314588B}}.

In the following, we discuss a few other implications of our model.

\subsection{Progenitor of EP240414a}\label{sec:progenitor}

{The initial volume-equivalent radius of the cocoon $R_{\rm c0} = (3V_{\rm c0}^{\rm iso}/4\pi)^{1/3}$ is related to the radius of the progenitor star $R_{\star}$. The relation is affected by the lateral expansion of the two cocoon components.}

{ Since the inner cocoon stays laterally confined by the outer cocoon's pressure during the acceleration phase and different parts of it lose causal contact after the expansion speed becomes relativistic, it is reasonable to expect that the opening angle of the inner cocoon stays roughly constant during the entire history of expansion. Thus, from the inner cocoon cooling emission, we infer the stellar radius to be $R_*\approx R_{\rm ic0} \simeq 3R_\odot$, which is only weakly affected by this choice \citep[see][for a discussion including the initial and final opening angles of the inner cocoon]{Nakar2017ApJ...834...28N}. }


{ The outer cocoon undergoes significant lateral expansion and as it eventually becomes quasi-spherical. If the opening angle of the outer cocoon at breakout is $\theta_{\rm oc0}$, then we have $R_{\star} \simeq 3R_{\rm oc0}(\theta_{\rm oc0}/15^{\circ})^{-2/3}$, because the initial volume is $V_{\rm oc0} = 2\pi \theta_{\rm oc0}^2 R_\star^3/3$ (which is equal to $V_{\rm oc0}^{\rm iso}$ as the outer cocoon is quasi-spherical when observed). In this picture, we find that, for an initial cocoon opening angle of $\theta_{\rm oc0}=15^{\rm o}$, the inferred progenitor's radius from the outer cocoon $R_\star \simeq 15.0^{+16.5}_{-11.4}R_\odot$ for an optimal color correction factor $f_{\tau} = 1.5$. The stellar radius inferred from the outer cocoon emission is larger than that from the inner cocoon, but the discrepancy is not statistically significant (only at the $1\sigma$ level). Due to the large theoretical and observational uncertainties for the outer cocoon emission (see \S \ref{sec:phase2_outer_cocoon}), we do not rule out the current model based on this weak discrepancy. If future observations confirm this discrepancy, additional heating of the outer cocoon after the breakout may be needed. This is beyond the scope of the current paper.
}

{ We conclude that the inner cocoon cooling emission constrains the progenitor's radius to be around $R_\star\sim 3R_\odot$. This is larger than the pre-SN radius of a massive helium star or a carbon-oxygen star from current stellar evolution calculations \citep[typically $\lesssim 1\,R_\odot$; see][for a review]{Levan2016SSRv..202...33L}. This discrepancy is difficult to reconcile given that the constraint from the inner cocoon emission is robust in our model. }

In fact, the radii of LGRB progenitors are poorly constrained by observations. It is possible that EP240414a had a progenitor star distinct from those of typical LGRBs, as also suggested by \cite{Sun2024arXiv241002315S} based on the unusual galactic environment. Alternatively, the radii of LGRB progenitors may be larger than the expectation from our current understanding of stellar evolution. For instance, the progenitor of EP240414a may have a low-mass extended envelope \citep[as suggested by][]{Hamidani2025_2025arXiv250316243H}.



Another possibility is that, if the progenitor is a helium star in a binary system \citep[e.g.][]{Izzard2004MNRAS.348.1215I,Bogomazov2009ARep...53..325B}, the outer boundary of the progenitor could correspond to the Roche lobe. For example, the progenitor could be a tidally spun-up helium star with a mass of $M_\star\sim5\,M_\odot$ in a close binary with an orbital period of $P_{\rm orb}\sim1\,{\rm day}$ \citep{fuller22_tidal_spin_up,hu2023}. If the helium star fills its Roche lobe before explosion, its radius is given by the Roche radius \citep{1983ApJ...268..368E}
\begin{equation}
    R_* = R_{\rm L}\approx4R_{\odot}\,(M_\star/5\,M_\odot)^{1/3}(P_{\rm orb}/1\,{\rm day})^{2/3}.
\end{equation}
However, in this case, the physical reason for the Roche lobe overflow is unclear --- heating of the outer envelope by internal gravity waves is possible \citep[e.g.,][]{2012MNRAS.423L..92Q, 2017MNRAS.470.1642F, 2021ApJ...906....3W}.
Furthermore, if the progenitor of EP240414a is indeed a helium star, the absence of helium in the spectra of the associated SN Ic-BL suggests that helium is likely hidden \citep{Piro_2014ApJ...792L..11P, Branch2017}. 

We also note that EP240414a has an unusually large offset of 26.3\,kpc from the center of its potential host galaxy, which is an unprecedented feature among SNe Ic-BL and GRB-SNe \citep{Sun2024arXiv241002315S}. This may indicate the presence of a hidden star formation region in the outskirts of the host galaxy or a faint satellite galaxy at the location of EP240414a. 
Searching for star formation regions in this area by the Hubble/James Webb Space Telescope will be crucial to understand the peculiar offset and to test models.

\subsection{Other Viewing Angles}
\label{sec:angle}

\begin{figure}
    \centering
    \includegraphics[width=1.0\linewidth]{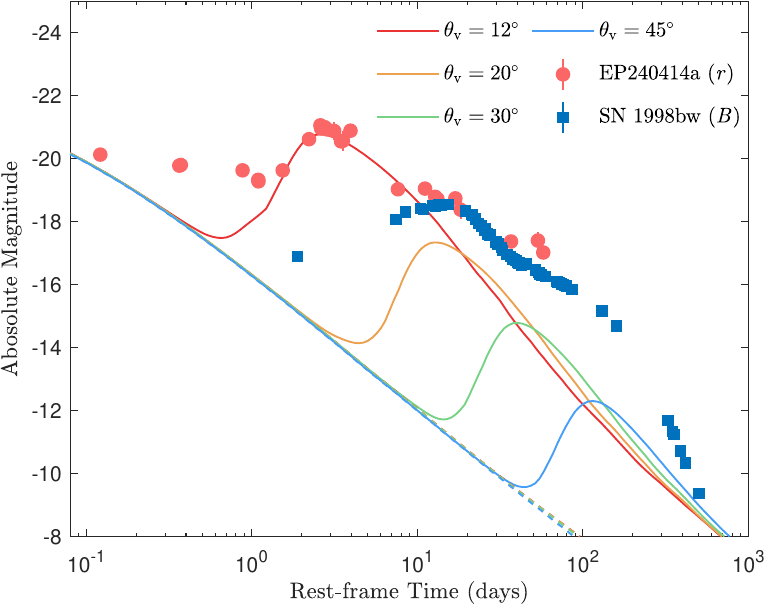}
    \includegraphics[width=1.0\linewidth]{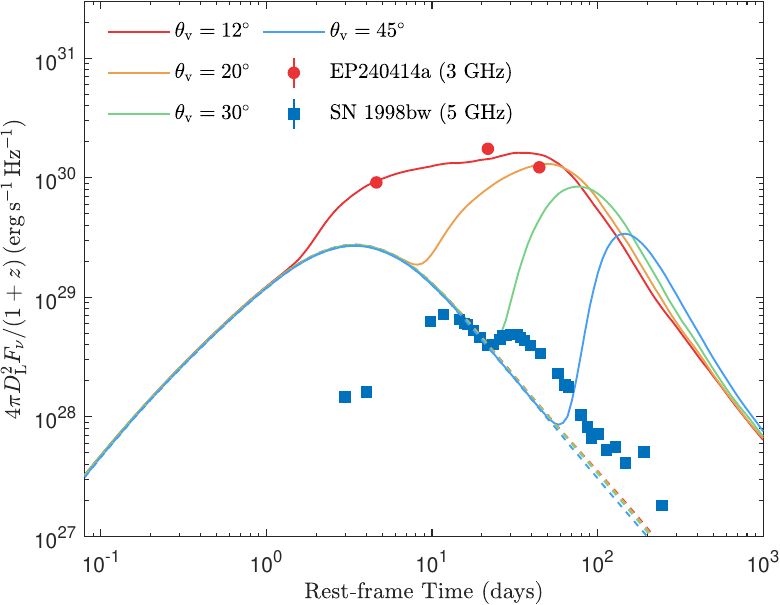}
    \caption{Afterglow emission at various viewing angles, $\theta_{\rm v} = 12^{\circ}$ (red), $20^{\circ}$ (orange), $30^{\circ}$ (green) and $45^{\circ}$ (blue). Solid lines include the contributions from the jet and inner cocoon, and dashed lines are for the inner cocoon only.  In the upper panel, the red circles and blue squares represent $r$-band data of EP240414a and $B$-band data of SN\,1998bw \citep{Clocchiatti2011AJ....141..163C}, and the model predictions are for $r$-band at $z=0.4$. In the lower panel, the red circles and blue squares mark the radio lightcurves of EP240414a at 3 GHz and SN\,1998bw at 5 GHz \citep{Kulkarni1998Natur.395..663K}, and the model predictions are for observer's frequency of 3 GHz at $z=0.4$, corresponding to $\sim$5 GHz in the rest-frame.
        }
    \label{fig:angle}
\end{figure}

The appearances of the jet-cocoon system with the same physical parameters but for different viewing angles are illustrated in Figure \ref{fig:angle}. We consider the optical and radio afterglows at $\theta_{\rm v}=20^{\circ},\,30^{\circ},$ and $45^{\circ}$ and compare them with the SN\,1998bw/GRB\,980625 data \citep{Kulkarni1998Natur.395..663K,Clocchiatti2011AJ....141..163C}.  The radio afterglows are bright and detectable at these large viewing angles. In contrast, the optical afterglows can easily be buried by the SN emission for 
$\theta_{\rm v}\gtrsim20^{\circ}$, which explains why the optical bump in EP240414a has rarely been 
observed in previous studies. The observation window for optical bumps is only accessible to observers within $\theta_{\rm j0}\lesssim\theta_{\rm v}\lesssim20^{\circ}$. To constrain the jet fraction in SNe Ic-BL, we encourage a systematic search for early-time optical ``red excess'' in existing/future datasets. 

\subsection{Detection Rate}
The event rate density of SN Ibc in the local universe ($z<0.1$) is $\mathcal{R}_{\rm Ibc}\approx 2.6\times10^{4}\,{\rm Gpc}^{-3}{\rm yr}^{-1}$ \citep{Li2011MNRAS.412.1473L}. 
Given that SNe Ic-BL are only a small portion ($\sim\!5\%$) of SNe Ibc \citep{Shivvers2017PASP..129e4201S}, the event rate density should be $\mathcal{R}_{\rm Ic\mbox{-}BL}\sim1.3\times10^{3}\,{\rm Gpc}^{-3}{\rm yr}^{-1}$.
EP240414a implies a limited observing window with a beaming factor $f_{\rm b}\approx0.02$, so the observable rate of EP240414a-like events in the local universe is $\mathcal{R}_{\rm EP240414a}\sim 20f_{\rm jet}\,{\rm Gpc}^{-3}{\rm yr}^{-1}$, where $f_{\rm jet}$ is the fraction of SNe Ic-BL with successful jets similar to EP240414a.
Ignoring redshift evolution, we infer the all-sky rate of EP240414a-like events to be $\sim\!73 f_{\rm jet}\,{\rm yr}^{-1}$ within $z=0.4$. Since EP-WXT covers $\sim\!8.7\%$ of the sky with an operational duty cycle of $50\%$ \citep{Yuan2022hxga.book...86Y,Sun2024arXiv241002315S},
one can estimate the detection rate of E240414a-like sources by the EP to be $3f_{\rm jet}\,{\rm yr}^{-1}$ within $z=0.4$. The detection of EP240414a shortly after the launch of EP constrains the fraction of SNe Ic-BL with successful jets to be $f_{\rm jet}\gtrsim 1\%$ (at 95$\%$ confidence level allowing for Poisson errors).

On the other hand, by comparing EP240414a with 36 known SNe Ic-BL, we find that its radio emission is the most luminous among the entire sample \citep{Corsi2016ApJ...830...42C,Corsi2023ApJ...953..179C}, which constrains the upper limit of the jet fraction to be $f_{\rm jet}\lesssim 10\%$ (95$\%$ confidence level). We also find that these two constraints are in tension at the 68$\%$ confidence level.

Therefore, we estimate the jet fraction to be $1\%\lesssim f_{\rm jet}\lesssim 10\%$, perhaps due to weaker jets being choked by the stellar envelope \citep[e.g.][]{meszaros2001PhRvL..87q1102M}.


\subsection{Predictions}
\label{sec:pre}
Our model gives several testable predictions:

(1) The 1.5 GHz radio afterglow of EP240141a will remain detectable by VLA and Giant Meterwave Radio Telescope \citep{Perley2011ApJ...739L...1P,Patra2019MNRAS.483.3007P} up to 1 yr after the trigger. This is because the radio emission at low frequencies ($0.8-1.5$\,GHz) becomes optically thin around 100\,d. The 5-sigma sensitivity of the VLA for a 3-hr exposure is shown in Figure \ref{afterglow2}.

(2) Afterglow emission may re-emerge in the optical band at $t\gtrsim(1+z)\,500$\,d. Comparing the late-time optical lightcurve of SN\,1998bw with theoretical off-axis afterglow in Figure \ref{fig:angle}, we find that the afterglow will eventually exceed the SN nebular emission as the latter decays exponentially. However, we note that SN\,1998bw did not have a successful ultra-relativistic GRB jet, so the purpose here is to demonstrate the SN nebular emission. For other nearby events with successful jets, the late-time afterglow emission may be detectable a few years after the discovery.

(3) In a small fraction of SNe Ic-BL (of the order 1\%, those with successful jets and viewed at $\theta_{\rm v}\lesssim 20^\circ$), we expect to detect a ``red excess'' within the first 3 to 10 days due to the orphan afterglow produced by an off-axis jet. This can be tested by future observations with the Very Rubin Observatory.

(4) Prior to the jet prompt emission and cocoon cooling emission, the shock breakout emission may arise as a precursor signal. Relativistic shock breakout emission is less luminous ($L\lesssim10^{46}{\rm erg\,s^{-1}}$), much hotter ($kT\gtrsim50$ keV), much shorter lived ($\lesssim 10\rm\,s$)\citep{Nakar2012ApJ...747...88N,Nakar2017ApJ...834...28N} compared to cocoon cooling emission.  Non-relativistic shock breakout emission (at larger viewing angles) can be identified by their lower luminosities, lower temperatures, and longer durations. The shock breakout signal can be used to provide independent constraints on the radius of the progenitor star.

\begin{acknowledgments}
The authors thank Yuhan Yao, { Paz Beniamini}, Xiang-Yu Wang, Jia Ren, Yu-Jia Wei, Yun-Wei Yu, Ilya Mandel, and Liang-Duan Liu for useful discussions. JHZ's work is supported by the National Natural Science Foundation of China (grant numbers 124B2057, 13001075). 
\end{acknowledgments}

\appendix
\section{A General Effective Temperature Formulation}
\label{appendix}
We derive a general effective temperature formulation for the cocoon applicable to both relativistic and non-relativistic regimes. To include relativistic effects, one needs to consider the physical quantities in the co-moving frame. The isotropic thermal energy in the co-moving frame is $E'^{\rm iso }_{\rm th,c}=u^{\prime}_{\rm c}V^{\prime}_{\rm c}$, where $u^{\prime}_{\rm c}$ is the energy density and 
$V'^{\rm iso}_{\rm c}\approx4\pi R_{\rm c}^2\times R_{\rm c}/(2\Gamma_{\rm c})$ is the volume. Adopting the Lorentz transformation for energy, i.e., $E^{\rm iso}_{\rm th,c}=\Gamma_{\rm c}E'^{\rm iso}_{\rm th,c}$, 
the energy density can be expressed by $u^{\prime}_{\rm c}=E^{\rm iso}_{\rm th,c}/(2\pi R^3_{\rm c})$. Under LTE, the energy density $u_{\rm c}^\prime$ is related to the radiation temperature $T_{\rm c}{\prime}$ in the co-moving frame by $u^{\prime}_{\rm c}=a{T^{\prime}_{\rm c}}^4$, where $a$ is the radiation density constant. Taking the Lorentz transformation $T_{\rm c}=\Gamma_{\rm c}T^{\prime}_{\rm c}$, we obtain $aT^4_{\rm c}=\Gamma^4_{\rm c}E^{\rm iso}_{\rm c,th}/(2\pi R^3_{\rm c})$.
Importantly, $T_{\rm c}$ is not the observed temperature since the cocoon is optically thick ($\tau_{\rm c}>1$). Instead, the effective photospheric temperature 
is $T_{\rm c,eff}\approx T_{\rm c}\tau^{-1/4}_{\rm c}$. At the diffusion time, the optical depth is $\tau_{\rm c}(t_{\rm c,diff})=\beta^{-1}_{\rm c}$, which gives the effective temperature at peak luminosity
\begin{equation}
   \label{dyneq1}
   aT^4_{\rm c,eff}=\frac{\beta_{\rm c}\Gamma^4_{\rm c}E^{\rm iso}_{\rm c,th}(R_{\rm c,diff})}{2\pi R^3_{\rm c,diff}}.
\end{equation}
Since $R_{\rm c,diff}=\beta_{\rm c}c\,t_{\rm c,diff}/(1+\beta_{\rm c})$, $\Gamma^{-2}_{\rm c}=(1-\beta^2_{\rm c})=(1+\beta_{\rm c})(1-\beta_{\rm c})$, and $L_{\rm c,peak} \approx E^{\rm iso}_{\rm c,th}(R_{\rm c,diff})/t_{\rm c,diff}$, Equation (\ref{dyneq1}) can be  expressed further as
\begin{equation}
  \label{dyneq2}
  aT^4_{\rm c,eff}=\frac{(1-\beta_{\rm c})L_{\rm c,peak}}{2\pi(1+\beta_{\rm c})^2\beta^2_{\rm c} t^2_{\rm c,diff}c^3}.
\end{equation}
Here, $T_{\rm c,eff}$, $L_{\rm c,peak}$ ,and $t_{\rm c,diff}$ are observables if the gas-radiation mixture is in LTE ($\eta_0\leq1$). 
Equation (\ref{equ:temperature}) can be obtained using $a=4\sigma_{\rm SB}/c$. 

{
When the outer cocoon reaches the diffusion radius $R_{\rm c,diff}\sim 10^{14}\rm\,cm$, its density is $\rho_{\rm c}\sim 10^{-12}{\rm g\,cm^{-3}}M_{\rm c,-2}R^{-3}_{\rm c,diff,14}$. At such a low density and temperature $T_{\rm c}\gtrsim 10^{4.5}\rm\, K$, the opacity is dominated by electron scattering such that the Rosseland-mean opacity is $\kappa_{\rm R} \approx \kappa_{\rm s}\sim 0.1{\rm\, cm^{2}\, g^{-1}}$. The color temperature of the observed emission is higher than the effective temperature when the effective optical depth $\tau_{\rm eff}\simeq \sqrt{\tau_{\rm a}\tau_{\rm s}}$ (where $\tau_{\rm a}\simeq \rho_{\rm c}\kappa_{\rm a} R_{\rm c,diff}$ and $\tau_{\rm s}\simeq \rho_{\rm c} \kappa_{\rm s} R_{\rm c,diff}$ are the absorption and scattering optical depths) is much lower than the optical depth $\tau_{\rm oc}\approx \tau_{\rm s}\simeq 1/\beta_{\rm oc}$ used in the main text. The color temperature is determined by the layer where the exterior effective temperature is roughly one, and in the limit of $\kappa_{\rm s}\gg \kappa_{\rm a}$, we obtain the following color correction
\begin{equation}
    T_{\rm oc,peak} = f_{\rm \tau} T_{\rm oc,eff}, \ \ f_{\tau} \simeq \min\left[\tau_{\rm oc}^{1/4}, (\kappa_{\rm s}/\kappa_{\rm a})^{1/8}\right]
\end{equation}
The case of $\tau^{1/4}_{\rm oc}$ is appropriate when the total $\tau_{\rm eff}$ is below unity whereas the other case of $(\kappa_{\rm s}/\kappa_{\rm a})^{1/8}$ is for the case for total $\tau_{\rm eff} > 1$. Since $\tau_{\rm oc}\approx 1/\beta_{\rm oc}$ at the diffusion radius, we find a maximum color correction factor of $\tau_{\rm oc}^{1/4}\approx1.5(\beta_{\rm oc}/0.2)^{-1/4}$.
}

\section{Methods of Afterglow Computations}
\label{appendix2}
Here, we will briefly describe the  afterglow model we adopted in this paper, which follows \cite{Zheng2024ApJ...966..141Z}.

In our model, the dynamics of the bulk Lorentz factor $\Gamma$ at radius $R$ is given by \citep{Nava2013MNRAS.433.2107N} 
\begin{equation}
\frac{{d\Gamma }}{{dR}} =  - \frac{{4\pi {R^2}\rho {c^2}\Gamma \left( {{\Gamma ^2} - 1} \right)\left( {\hat \gamma \Gamma  - \hat \gamma  + 1} \right) - \left( {\hat \gamma  - 1} \right)\Gamma \left( {\hat \gamma {\Gamma ^2} - \hat \gamma  + 1} \right)\left( {3U/R} \right)}}{{{\Gamma ^2}\left( {{m_0} + m_{\rm sw}} \right){c^2} + \left( {{{\hat \gamma }^2}{\Gamma ^2} - {{\hat \gamma }^2} + 3\hat \gamma  - 2} \right)U}},
\end{equation}
where $m_0=f_{\rm \{j,c\}}(\theta_{\rm \{j0,c0\}})E_{\rm \{j,c\},k}^{\rm iso}/(\Gamma_0-1)c^2$ is the rest-mass of the jet/cocoon, $m_{\rm sw}=\int4\pi R^2f_{\rm \{j,c\}}(\theta_{\rm \{j,c\}})\rho(R)dR$ is the swept-up rest-mass, $\rho(R)=m_{\rm p}n(R)$ is the density of the ambient environment with $m_{\rm p}$ being the proton mass, $f_{\rm \{j,c\}}(\theta_{\rm \{j,c\}})=(1-\cos\theta_{\rm \{j,c\}})/2$ is the beaming factor, $\hat{\gamma}$ is the adiabatic index of the shock-heated gas, and $U$ is the internal energy of the shock-heated gas. We evolve the internal energy with radius by
\begin{equation}
\frac{{dU}}{{dR}} = 4\pi {R^2}\rho \left( {1 - \varepsilon } \right)\left( {\Gamma  - 1} \right){c^2} - \left( {\hat \gamma  - 1} \right)\left( {\frac{3}{R} - \frac{1}{\Gamma }\frac{{d\Gamma }}{{dR}}} \right)U,
\end{equation}
where $\varepsilon$ is the radiation efficiency of the forward shock. 
We adopt the same expressions for radiation efficiency $\varepsilon$ and adiabatic index $\hat{\gamma}$ as those used in \cite{Zheng2024ApJ...966..141Z}. The lateral expansion of the shocked gas is described by \citep{Huang2000ApJ...543...90H}
\begin{equation}
    \frac{{d\theta_{\rm \{j,c\}}}}{{dR}} = \frac{{{c'_{\rm s}}}}{\Gamma\beta Rc}, \qquad
    {c'}_{\rm s}^2 = \left[ {\frac{{\hat \gamma \left( {\hat \gamma  - 1} \right)\left( {\Gamma  - 1} \right)}}{{1 + \hat \gamma \left( {\Gamma  - 1} \right)}}} \right]{c^2},
\end{equation}
where $c'_{\rm s}$ is the speed of sound in the comoving frame, $\beta=\sqrt{1-\Gamma^{-2}}$ is the velocity of shocked gas in the unit of $c$. 
Using the above equations with appropriate initial conditions, one can obtain the evolution of the bulk Lorentz factor $\Gamma$, the shock radius $R$ and the angle $\theta_{\rm \{j,c\}}$ of the jet and cocoon. We then consider synchrotron cooling, as well as inverse Compton cooling, which includes the Klein-Nishina effect, to model the electron spectra and calculate the spectral power $P'_\nu$ in the co-moving frame of the shocked gas \citep[see][for more details]{Zheng2024ApJ...966..141Z}.

To obtain the observed fluxes along different lines of sight, one needs to consider the effect of equal arrival time surfaces. The observer's time is obtained by 
\begin{equation}
    {t_{\rm obs}} = \left({1 + z}\right)\int {\frac{{1 - \beta \cos \chi }}{{\beta c}}dR},
\end{equation}
where $\cos\chi =\cos{\theta_{\rm v} }\cos\theta  + \sin{\theta_{\rm v}}\sin\theta\cos\varphi$. 
The observed flux density can be obtained by integrating the equal arrival time surface
\begin{equation}
    {F_\nu } = \frac{{1 + z}}{{4\pi D_L^2}}\iint {{\mathcal{D}^3}\frac{{{{P}_\nu' }}}{{4\pi }}d\Omega_{\rm j} },
\end{equation}
where $D_{\rm L}$ is the luminosity distance and $\mathcal{D}\equiv[\Gamma(1-\beta\cos\chi)]^{-1}$ is the Doppler factor. 
The integration of solid angle $\Omega_{\rm j}$ is performed over the emitting region from the center of explosion.
The frequency of the photons in the observer's frame is thus given by $\nu_{\rm obs}=\mathcal{D}\nu^{\prime}/(1+z)$.

\section{Constraining the jet Lorentz factor by the lack of gamma-ray prompt emission}
\label{appendix3}

\begin{figure*}
    \centering
    \includegraphics[scale=0.5]{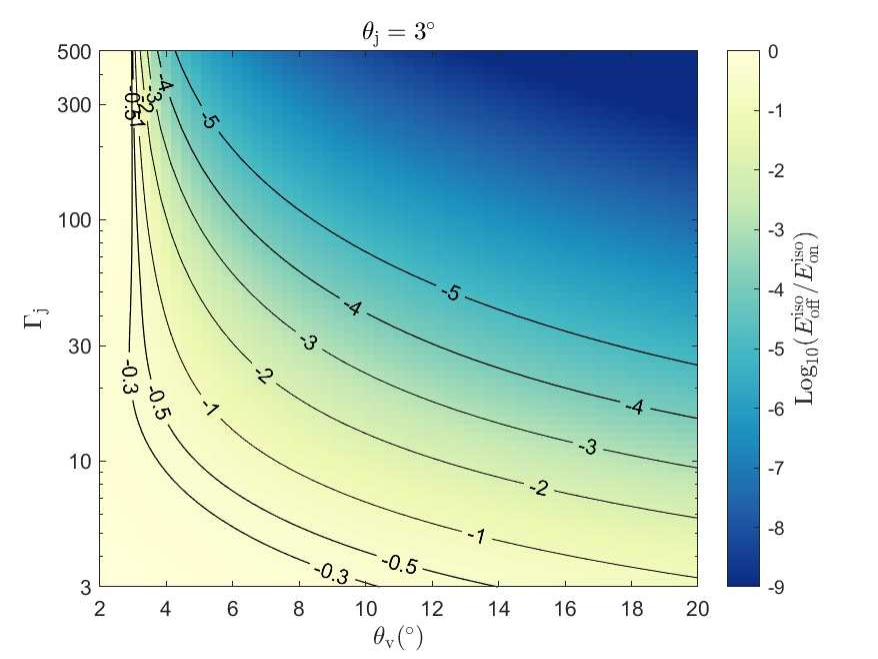}
    \includegraphics[scale=0.5]{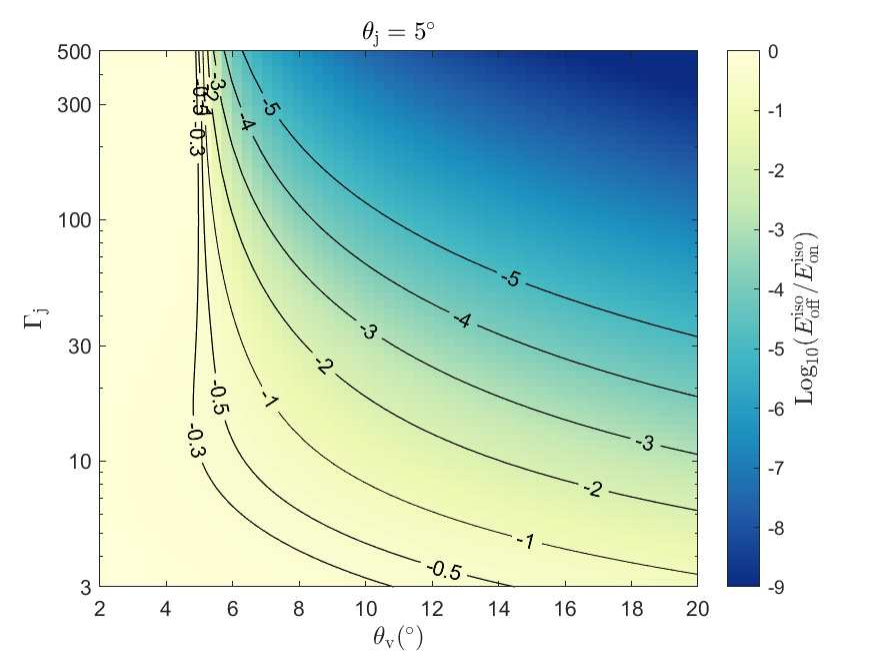}
    \caption{The logarithmic isotropic equivalent energy ratio between off-axis observers and on-axis observers (the color map) with respect to the jet Lorentz factor and viewing angle. The left and right panels are for jet half opening angle $\theta_{\rm j}=3^{\circ}$ and $\theta_{\rm j}=5^{\circ}$, respectively. Contour lines are energy ratios corresponding to specific values ranging from $10^{-0.3}$ to $10^{-5}$.
    }
    \label{app:gamma}
\end{figure*}
The prompt emission from the jet is strongly de-boosted when the observer is far from the jet axis ($\theta_{\rm v}>\theta_{\rm j}$). To quantify this effect we calculate the observed isotropic equivalent gamma-ray energy ratio ($E^{\rm iso}_{\rm \gamma, off}/E^{\rm iso}_{\rm \gamma, on}$) between off-axis and on-axis observers. 

 For simplicity, we assume at top-hat jet and that the specific intensity of the prompt gamma-ray emission in the comoving frame $I'_{\nu}$ is constant over the duration $T'$. Applying the Lorentz transformation $I_{\nu}=\mathcal{D}^3(\theta_{\rm v},\theta_{\rm },\varphi)I'_{\nu}$, we obtain the isotropic equivalent specific luminosity is $L_{\nu}=4\pi R^2\iint\mathcal{D}^3(\theta_{\rm v},\theta_{\rm },\varphi)I'_{\nu}d\Omega_{\rm j}$. Considering the transformation of frequency $\nu=\mathcal{D}(\theta_{\rm v},\theta_{\rm },\varphi)\nu'$ and time $T=T'/\mathcal{D}(\theta_{\rm v},\theta_{\rm },\varphi)$, the isotropic energy in the observer's frame is $E^{\rm iso}_{\gamma}\sim \nu L_{\nu}T=4\pi R^2\iint\mathcal{D}^3(\theta_{\rm v},\theta_{\rm },\varphi)\nu' I'_{\nu}T'd\Omega_{\rm j}$. We neglect the cosmological effects which do not affect the ratio. Therefore, the isotropic energy ratio between off-axis (at the given $\theta_{\rm v}$) and on-axis (at $\theta_{\rm v}=0$) observers is given by
\begin{equation}
    \frac{E^{\rm iso}_{\rm\gamma, off}}{E^{\rm iso}_{\rm \gamma, on}}=\frac{\iint\mathcal{D}^3(\theta_{\rm v},\theta_{\rm },\varphi)d\Omega_{\rm j}}{\iint\mathcal{D}^3(0,\theta_{\rm },\varphi)d\Omega_{\rm j}}.
\end{equation}
The ratio depends on the jet Lorentz factor $\Gamma_{\rm j}$, viewing angle $\theta_{\rm v}$ and jet half-opening angle $\theta_{\rm j}$. We present the logarithmic ratio of a top-hat jet on Figure \ref{app:gamma} with $\theta_{\rm j}=3^{\circ}$ and $\theta_{\rm j}=5^{\circ}$. The ratio is nearly a constant for observers at $\theta_{\rm v}<\theta_{\rm j}$ and declines rapidly as $E^{\rm iso}_{\rm \gamma, off}/E^{\rm iso}_{\rm \gamma, on}\propto\theta^{-6}_{\rm v}$ for off-axis observers ($\theta_{\rm v}>\theta_{\rm j}$). Given that the typical isotropic energy for GRB prompt emissions is $E_{\gamma,\rm on}^{\rm iso}\sim10^{51}$ erg,  relativistic beaming would lead to $E^{\rm iso}_{\rm \gamma,off}<10^{48}{\rm erg}$ for $\theta_{\rm v}\geq 10^{\circ}$ if $\Gamma_{\rm j}\gtrsim 30$, leading to the non-detection of EP and other X-ray satellites.  If the jet has an angular structured profile, a higher Lorentz factor is required for the same ratio $E^{\rm iso}_{\rm \gamma, off}/E^{\rm iso}_{\rm \gamma, on}$ and viewing angle.

\bibliography{reference}{}
\bibliographystyle{aasjournal}

\end{document}